\documentclass[
12pt, % The default document font size, options: 10pt, 11pt, 12pt
oneside, % Two side (alternating margins) for binding by default, uncomment to switch to one side
english, % ngerman for German
singlespacing, % Single line spacing, alternatives: onehalfspacing or doublespacing
headsepline, % Uncomment to get a line under the header
]{MastersDoctoralThesis} % The class file specifying the document structure

\usepackage[utf8]{inputenc} % Required for inputting international characters
\usepackage[T1]{fontenc} % Output font encoding for international characters

\usepackage{mathpazo} % Use the Palatino font by default

\usepackage[autostyle=true]{csquotes} % Required to generate language-dependent quotes in the bibliography

\usepackage{tikz} % Use tiks to display state machine  diagram
\usetikzlibrary{arrows,automata,positioning, arrows.meta}
\tikzset{
->, % makes the edges directed
initial text=$ $, % sets the text that appears on the start arrow
}

\usepackage{longtable} % longtable to continue table on next page

\usepackage{mathrsfs} % math text font 

\usepackage{subcaption} % package to add child caption inside main caption parent

\usepackage{placeins}
\usepackage{listings} % package to style code listing
\usepackage{xcolor}
\definecolor{codegreen}{rgb}{0,0.6,0}
\definecolor{codegray}{rgb}{0.5,0.5,0.5}
\definecolor{codepurple}{rgb}{0.58,0,0.82}
\definecolor{backcolour}{rgb}{0.95,0.95,0.92}

\lstdefinestyle{mystyle}{
    backgroundcolor=\color{backcolour},   
    commentstyle=\color{codegreen},
    keywordstyle=\color{magenta},
    numberstyle=\tiny\color{codegray},
    stringstyle=\color{codepurple},
    basicstyle=\ttfamily\footnotesize,
    breakatwhitespace=false,         
    breaklines=true,                 
    captionpos=b,                    
    keepspaces=true,                 
    numbers=left,                    
    numbersep=5pt,                  
    showspaces=false,                
    showstringspaces=false,
    showtabs=false,                  
    tabsize=2
}

\thesistitle{Verifying User Interfaces using SPARK Ada: A case study of the T34 syringe driver} % Your thesis title, this is used in the title and abstract, print it elsewhere with \ttitle
\supervisor{Dr. Jens \textsc{Blanck}} % Your supervisor's name, this is used in the title page, print it elsewhere with \supname
\examiner{} % Your examiner's name, this is not currently used anywhere in the template, print it elsewhere with \examname
\degree{Master of Science} % Your degree name, this is used in the title page and abstract, print it elsewhere with \degreename
\author{Peterson \textsc{Jean} (2109887)} % Your name, this is used in the title page and abstract, print it elsewhere with \authorname
\addresses{} % Your address, this is not currently used anywhere in the template, print it elsewhere with \addressname

\subject{Computer Science} % Your subject area, this is not currently used anywhere in the template, print it elsewhere with \subjectname
 % Keywords for your thesis, this is not currently used anywhere in the template, print it elsewhere with \keywordnames
\university{\href{https://swansea.ac.uk}{Swansea University}} % Your university's name and URL, this is used in the title page and abstract, print it elsewhere with \univname
\department{\href{}{Department of Computer Science}} % Your department's name and URL, this is used in the title page and abstract, print it elsewhere with \deptname
\group{\href{}{}} % Your research group's name and URL, this is used in the title page, print it elsewhere with \groupname
\faculty{\href{http://faculty.university.com}{Faculty Name}} % Your faculty's name and URL, this is used in the title page and abstract, print it elsewhere with \facname

\AtBeginDocument{
\hypersetup{pdftitle=\ttitle} % Set the PDF's title to your title
\hypersetup{pdfauthor=\authorname} % Set the PDF's author to your name
\hypersetup{pdfkeywords=\keywordnames} % Set the PDF's keywords to your keywords
\hypersetup{colorlinks=false} % having obvious links in the PDF but not the printed text,
\hypersetup{hidelinks}
}

\begin{document}

\frontmatter % Use roman page numbering style (i, ii, iii, iv...) for the pre-content pages

\pagestyle{plain} % Default to the plain heading style until the thesis style is called for the body content

%----------------------------------------------------------------------------------------
%	TITLE PAGE
%----------------------------------------------------------------------------------------

\begin{titlepage}
\begin{center}

% \vspace*{.06\textheight}
% {\scshape\LARGE \univname\par}\vspace{1.5cm} % University name
\textsc{\Large }\\[0.5cm] % Thesis type

\includegraphics[scale= 0.5]{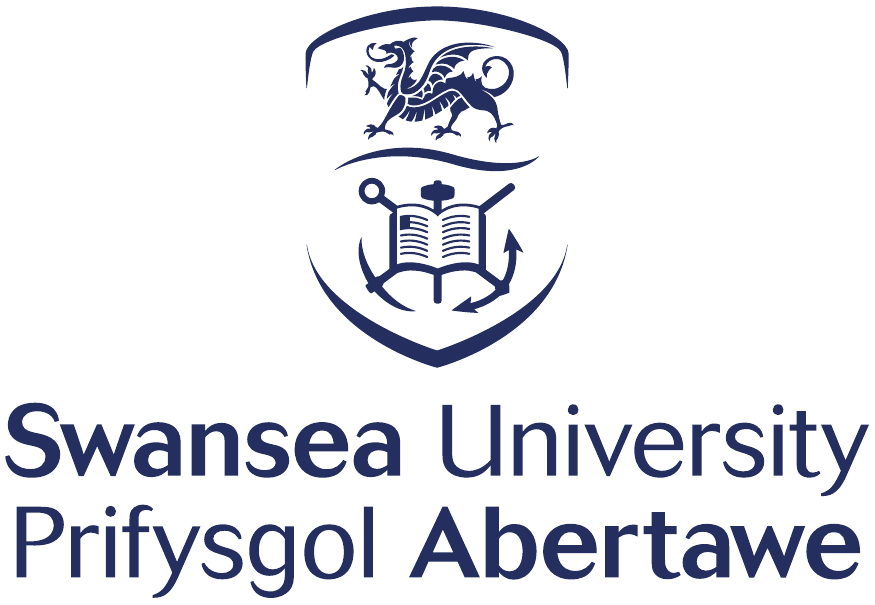} % University/department logo - uncomment to place it
\HRule \\[0.4cm] % Horizontal line
{\huge \bfseries \ttitle\par}\vspace{0.4cm} % Thesis title
\HRule \\[1.5cm] % Horizontal line

\large
\emph{}\\
\href{http://www.johnsmith.com}{\authorname} % Author name - remove the \href bracket to remove the link 

% \begin{minipage}[t]{0.4\textwidth}
% \begin{flushright} \large
% \emph{Supervisor:} \\
% \supname % Supervisor name - remove the \href bracket to remove the link  
% \end{flushright}
% \end{minipage}\\[3cm]
 
\vfill

\large \textit{Project Dissertation submitted to Swansea University in Partial Fulfilment \\ for the Degree of  \degreename}\\[0.3cm] % University requirement text
\textit{in the}\\[0.4cm]
\groupname \deptname\\[2cm] % Research group name and department name
 
\vfill

{\large September 30, 2022}\\[4cm] % Date

\vfill
\end{center}
\end{titlepage}

%----------------------------------------------------------------------------------------

%----------------------------------------------------------------------------------------
%	QUOTATION PAGE
%----------------------------------------------------------------------------------------

% \vspace*{0.2\textheight}

% \noindent\enquote{\itshape TODO: AknolowedgmentThanks to my solid academic training, today I can write hundreds of words on virtually any topic without possessing a shred of information, which is how I got a good job in journalism.}\bigbreak

% \hfill Peterson Jean

%----------------------------------------------------------------------------------------
%	ABSTRACT PAGE
%----------------------------------------------------------------------------------------

\begin{abstract}
\addchaptertocentry{\abstractname} % Add the abstract to the table of contents
The increase in safety and critical systems improved Healthcare.  Due to their risk of harm, such systems are subject to stringent guidelines and compliances.  These safety measures ensure a seamless experience and mitigate the risk to end-users.  Institutions like the Food and Drug Administration or the NHS \cite{1400-1700_iec_nodate,noauthor_t34_nodate} respectively established international standards and competency frameworks to ensure industry compliance with these safety concerns.  Medical device manufacturing is mainly concerned with standards.  
Consequently, These standards now advocate for better human factors considered in user interaction for medical devices.  This forces manufacturers to rely on heavy testing and review to cover many of these factors during development.  Sadly, many human factor risks will not be caught until proper testing in real life, which might be catastrophic in the case of an ambulatory device like the T34 syringe pump.  Therefore, effort in formal methods research may propose new solutions in anticipating these errors in the early stages of development or even reducing their occurrence based on the use of standard generic model.  These generically developed models will provide a common framework for safety integration in industry and may potentially be proven using formal verification mathematical proofs.  This research uses SPARK Ada's formal verification tool against a behavioural model of the T34 syringe driver.  A Generic Infusion Pump \cite{noauthor_generic_nodate} model refinement is explored and implemented in SPARK  Ada.  As a subset of the Ada language, the verification level of the end prototype is evaluated using SPARK.  Exploring potential limitations defines the proposed model's implementation liability when considering abstraction and components of User Interface design in SPARK Ada.
\end{abstract}

%----------------------------------------------------------------------------------------
%	ACKNOWLEDGEMENTS
%----------------------------------------------------------------------------------------

\begin{acknowledgements}
\addchaptertocentry{\acknowledgementname} % Add the acknowledgements to the table of contents
   
 I would like to express my most profound appreciation to the faculty staff at Swansea University, who took the time to answer my queries.  Of course, I could not have undertaken this project without the help of my supervisor  Dr Jens Blanck who was always responsive to my queries.

 I am also grateful to the overall support of the Chevening Scholarships, the UK government’s global scholarship programme, funded by the Foreign, Commonwealth and Development Office (FCDO) and partner organisations.
 
A special thanks to my family and friends for believing in me even when I doubted myself throughout this journey

\end{acknowledgements}

%----------------------------------------------------------------------------------------
%	LIST OF CONTENTS/FIGURES/TABLES PAGES
%----------------------------------------------------------------------------------------

\tableofcontents % Prints the main table of contents

\listoffigures % Prints the list of figures

\listoftables % Prints the list of tables

%----------------------------------------------------------------------------------------
%	ABBREVIATIONS
%----------------------------------------------------------------------------------------

\begin{abbreviations}{ll} % Include a list of abbreviations (a table of two columns)

\textbf{PVS} & \textbf{P}rototype \textbf{V}erification \textbf{S}ystem\\
\textbf{GIP} & \textbf{G}eneric \textbf{I}nfusion \textbf{P}ump\\
\textbf{GIIP} & \textbf{G}eneric \textbf{I}nsulin \textbf{I}nfusion \textbf{P}ump\\
\textbf{GPCA} & \textbf{G}eneric \textbf{P}atient \textbf{C}ontrolled 
\textbf{A}nalgesic\\
{\textbf{SPARK}} & {\textbf{S}}PADE {\textbf{A}}da {\textbf{R}}atiocinative {\textbf{K}ernel}\\
\textbf{ISO} & \textbf{I}nternational \textbf{O}rganisation of \textbf{S}tandard \\
\textbf{POST} & \textbf{P}ower- \textbf{O}n \textbf{S}elf-  \textbf{T}est\\ 
\textbf{UI} &\textbf{U}ser \textbf{ I}nterface \\
\textbf{GUI} &\textbf{G}raphical \textbf{ U}ser \textbf{I}nterface\\
\textbf{GNOGA} &\textbf{G}NU\textbf{ O}mnificent \textbf{G}UI for \textbf{A}da\\
\end{abbreviations}

%----------------------------------------------------------------------------------------
%	PHYSICAL CONSTANTS/OTHER DEFINITIONS
%----------------------------------------------------------------------------------------

% \begin{constants}{lr@{${}={}$}l} % The list of physical constants is a three column table

% % The \SI{}{} command is provided by the siunitx package, see its documentation for instructions on how to use it

% Speed of Light & $c_{0}$ & \SI{2.99792458e8}{\meter\per\second} (exact)\\
% %Constant Name & $Symbol$ & $Constant Value$ with units\\

% \end{constants}

%----------------------------------------------------------------------------------------
%	SYMBOLS
%----------------------------------------------------------------------------------------

\begin{symbols}{lll} % Include a list of Symbols (a three column table)

$V$ &Volume & ml \\
%Symbol & Name & Unit \\

\addlinespace % Gap to separate the Roman symbols from the Greek

\begin{math}
\mathscr{A} 
\end{math} & Label Transition System &  \\

\end{symbols}

%----------------------------------------------------------------------------------------
%	DEDICATION
%----------------------------------------------------------------------------------------

% \dedicatory{For/Dedicated to/To my\ldots} 

%----------------------------------------------------------------------------------------
%	THESIS CONTENT - CHAPTERS
%----------------------------------------------------------------------------------------

\mainmatter % Begin numeric (1,2,3...) page numbering

\pagestyle{thesis} % Return the page headers back to the "thesis" style

% Include the chapters of the thesis as separate files from the Chapters folder
% Uncomment the lines as you write the chapters

% Chapter Template

\chapter{Introduction} % Main chapter title

\label{Chapter1} % Change X to a consecutive number; for referencing this chapter elsewhere, use \ref{ChapterX}

%----------------------------------------------------------------------------------------
%	SECTION 1
%----------------------------------------------------------------------------------------

\section{Problem and Aim}

Healthcare has evolved with the increase in safety and critical systems.  Despite their positive impact, these systems are subject to high risks.  Medical devices are subject to the most rigorous safety standards or requirements to ensure successful solutions and a seamless experience for end -users.  The medical device industry relies primarily on international standards such as IEC 62366-2, defined by the Food and Drug Administration\cite{1400-1700_iec_nodate}. This ISO standard provides a framework for leveraging human factors in user interaction for medical devices.  In addition, the Evaluation of Assurance Level 7 (EAL) standard \cite{noauthor_common_nodate} introduces crucial steps towards more formal proofs of device abstraction models to mitigate design flaws and mathematically prove expected system behaviour.

For many device manufacturers, interaction with user interfaces remains an intangible element limited to regulation rather than facilitating the solution delivery process.  After deployments, the user interface represents the only point of interaction for all user input and error checking.  Research on formal techniques has proven to be concise and accurate in analysing system design and identifying violations of requirement constraints in the design aspects of devices \cite{harrison_verification_2017}. The traditional approach to development tends to rely on testing to catch critical errors such as execution and overflow that may be difficult to detect in real life's early and critical stages.  This traditional approach is called "pre-market review".  The risks associated with this approach are higher, as the device is used by a wider audience and may lead to potential problems that the Pre-Market Review tests did not detect.  These potential problems are often detected in post-market surveillance, which relies on the oversight of accredited regulatory bodies such as the FDA\cite{masci_verification_2013}.

Infusion pumps are one of the most common safety-related medical devices, and their diverse implementation covers a wide range of applications.  To ensure compliance with a safety risk, they are classified as Medical Class II devices in the FDA's three main medical device risk classes \cite{masci_verification_2013}.   In the UK, previous critical incidents have led to the creation of a specific vigilance system for reporting incidents involving specific medical devices such as infusion pumps, as outlined in the DSVG \cite{noauthor_medical_nodate}. Other human factors incidents caused by inappropriate user interface design are also reported by the FDA   \cite{health_examples_2019} . In some cases, pump warnings are overused and displayed so often that they become a nuisance to the user, who usually tends to ignore them, which may lead to misuse or unexpected behaviour of the device.  Another typical case is the lack of clarity of warning messages or the absence of appropriate user-friendly information, which can make the learning curve difficult or lead to patient harm.  These design errors result from the challenging physical characteristics of the infusion pump, such as minimal screen displays or input interfaces, which makes it essential to ensure the safety and verification of user inputs and outputs.

The T34 syringe is a widely used pump in ambulatory services.  Considering the NHS competency framework \cite{noauthor_t34_nodate}, the variety of causes for each problem on the T34 does not effectively alleviate the workload of carers, as the decision space for the solution is highly dependent on their judgements.  For example, the possible cause "Pump defective" corresponds to 3 different problems in the troubleshooting space.  Although incident reports work as excellent post-market surveillance, it is more critical to prevent these incidents earlier in the design stages, as supported by formal verification methods.  In response, international initiatives such as the CHI+MED-supported Generic Infusion Pump Research Project are seeking innovative safety models to provide a more general and standardised identification of the risks associated with medical infusion pumps \cite{noauthor_generic_nodate}. The two models provided aim to demonstrate measures and techniques to improve safety by design and apply to any equipment.  They allow manufacturers to shorten the lengthy safety validation process by evaluating their pump design before the final regulatory review.  They address the initial concern of the industry about the cost and development time associated with formal product verification. 
 
Therefore, this research will focus on using refinement techniques of the generic infusion pump design model to identify safety risks in the design of the T34 syringe pump.  This refinement also reflects the behavioural model of the device.  The focus will be on the syringe control system to limit the scope of the implementation.  Can the proposed verified user interface model limit the error space of the T34 and report fewer ambiguous errors?  Tools such as Spark's Ada verification tool, a subset of the Ada language, offer a crucial advantage as it incorporates automatic provers and executable generation support.  It will help demonstrate verification of the refined T34 design model and specification.  The development of the user interface design specification will represent a model of the medical device.  As part of the refinement specification, the derived behavioural models will represent the user interface interactions in the software implementation and must be proven.  The software implementation of the model must be fully implemented in Spark Ada.  It should effectively simulate the interactions of an actual T34 pump device and possibly avoid run-time and overflow errors. 

Furthermore, a successful prototype may serve as support material for the training effort in an initiative such as the Learnpro T34 module of the CME \cite{noauthor_t34_nodate}

%----------------------------------------------------------------------------------------
%	SECTION 2
%----------------------------------------------------------------------------------------

\section{Roadmap}
This dissertation is divided into chapters, each representing a substantial section of the evolution of this project.  Chapter 2 presents a review of the literature on formal verification, an overview of the SPARK verifier and its tools and a consideration of the T34 syringe driver.  In Chapter 3, we discuss the ethical issues, the proposed architecture, the model design and the verification process in SPARK while considering the risks and safety requirements.  Chapter 4 covers an analysis of the results and some of their implications.  Finally, Chapter 5 discusses general thoughts on the work and the resulting limitations.  Furthermore, the potential future implications are explored. 

% Chapter Template

\chapter{Background and Literature Review} % Main chapter title

\label{Chapter2} % Change X to a consecutive number; for referencing this chapter elsewhere, use \ref{ChapterX}

%----------------------------------------------------------------------------------------
%	SECTION 1
%----------------------------------------------------------------------------------------

\section{Formal Methods for User Interface}

Several formal theories of user interface verification have promising use in academia and industry. Initially, formal methods were considered unrealistic because of their overhead in critical software development\cite{king_is_2000}. Gradually, remarkable results in industrial-scale systems have changed the perspective. Initially, researchers relied on the model checker for formal verification. However, in the medical device industry, the reinforcement of safety regulations has contributed to the emergence of more quality-oriented software concepts, such as model-based engineering \cite{masci_model-based_2013, zhang_hazard_2010} . Model-based engineering \cite{kampfner_model-based_1997} aims to develop safe and reliable medical devices by eliminating early design errors in the various system models. It focuses on research to create generic safety models for infusion pumps, such as the Generic Insulin Infusion Pump (GIIP) model and the Generic Patient Controlled Analgesia Infusion Pump Model \cite{noauthor_generic_nodate}. As part of the Generic Infusion Pump Research Project, these models aimed to provide a realistic generic user interface test bed for industry and research to collectively incorporate the latest research on the safety of these medical devices. Currently, two significant challenges prevent these projects from adopting formal verification widely. Firstly, some companies have kept the development of their devices proprietary. A second challenge is that few implementations exist for the existing generic model, mainly in the PVS verification system \cite{masci_model-based_2013}.

\subsection{PVS Verification}
In their PVS implementation, Harold et al.'s research on interface software is based on Rushby configuration diagrams to verify the consistency of a system \cite{masci_formal_2014}. They introduce an unconventional formal approach by verifying a source code of the actual implementation of the infusion pump software in a derived behavioural model. This model then verifies the satisfaction of the system's human factors. Their PVS Proof system analysis is convincing in detecting real interaction design problems for several manufacturers, even with a partial implementation of the software. This approach then requires human intervention. The researchers argue that the human intervention required in PVS is more beneficial to the analyst. However, it could represent a need for specific expertise that could add more subjectivity to the model. Their approach to generating test cases for their model could contribute to our research, given that our idealised T34 pump is a non-digital key-based device.

\subsection{Device Specification Verification}
Moreover, Harrison et al. used the simulation technique of Harold et al. in their description of the device specification model  \cite{harrison_verification_2017}. They presented an iterative methodology for verifying the usability requirements of a system. They first determine a complete, accurate snapshot of the device's user interface states. Compared to Harold et al., their technique offers the flexibility to use model checking or simulation techniques to validate the previously defined device model. They ensure that the defined model meets the standard FDA safety requirements for infusion pumps.

\subsection{Device Characteristics Verification}
Since manufacturers must inevitably comply with the safety requirements of the regulatory agency, researchers in \cite{ruksenas_developing_2014} also demonstrated that it is possible to verify user interfaces by tracing the requirements with simplified hierarchical classes that match the physical characteristics of their device. This method is called refinement and involves segmenting and proving the user interface requirements. According to the Pre-Market Review process, these proven segmentations appear to be sufficient evidence, i.e. comparison with other existing certified devices on the market. Despite its relevance, Rukšėnas et al. \cite{ruksenas_developing_2014} expressed concerns about the consistency of the different segments with each other. Overall, the two main advantages of this method are the possibility to verify concrete interfaces, either by proving that the manufacturer's user requirements are met or by reverse engineering an actual device. The other advantage is that it works well for abstracting non-numeric key-based input interfaces that match the physical characteristics of the T34 syringe pump driver.

\section{T34 Syringe Driver}

%-----------------------------------
%	SUBSECTION 1
%-----------------------------------
\subsection{Device Characteristics}
\label{Device Characterisitics}
% \begin{figure}[h]
% \centering
% \includegraphics[scale=0.5]{t34-parts}
% \caption{A T34 parts diagram }
% \label{fig:x t34-parts}
% \end{figure}

\begin{figure}[h]
\centering
\includegraphics[scale=0.4]{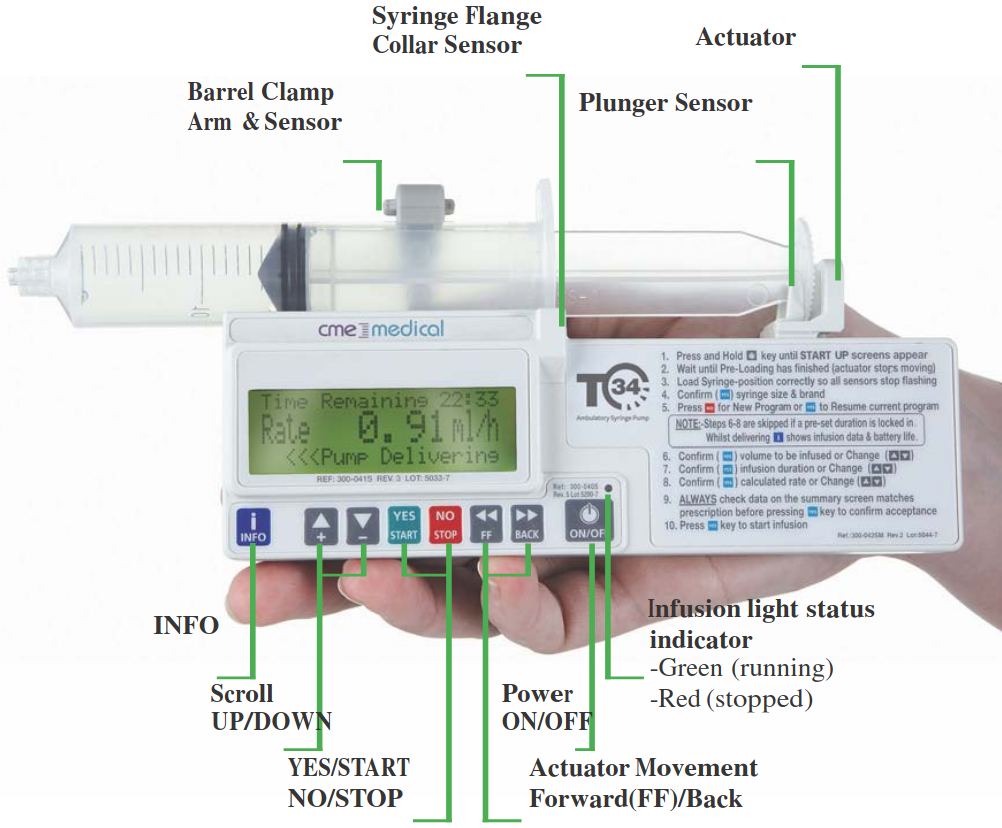}
\decoRule
\caption{A T34 parts diagram \cite{nhs_education_for_scotland_guidelines_nodate} }
\label{fig:T34}
\end{figure}
The physical characteristics of the T34 make it an ideal drug delivery device for ambulatory services, as it is lightweight, mobile and battery operated. These features make it ideal for long-term drug delivery, which is ideal in palliative care for subcutaneous drug delivery for 24 hours when the patient is unconscious, weak, or has other obstructions to the recommended oral drug delivery \cite{noauthor_t34_nodate}. Its long-term use for drug delivery also exposes it to an increased risk of patient harm, such as infection or device-related damage. As briefly discussed above, this risk justifies the need to classify it as a safety-critical device. The safety requirements and guidelines set by organisations such as the FDA for device manufacturers significantly minimise the imminent hazards and risks associated with devices. However, at the same time, non-device-related risks, such as infections due to patient mobility or skin reactions, are external harms that are almost impossible to take into account in the safety requirements of devices. Hence the justification for the importance of caring support personnel despite the device's autonomy for 24-hour administration.

The competency required to effectively and safely manage the T34 is considerably costly, both in human expertise and resources \cite{noauthor_t34_nodate}. For example, the device is regularly inspected and tested by the Medical Physics Department for compliance with safe use. However, nurses still need to be trained in part for immediate response according to updated guidelines on recent device incidents. There is likely a correlation between the frequency of use and the level of proficiency with the device, representing an additional external risk to the existing known error space \cite{noauthor_medical_nodate} for the T34 syringe driver. 
\subsection{Device UI Design}
The device's design addresses some of the common human factor issues, which is a great advantage over other infusion pump designs. Harold and Cairns \cite{noauthor_chimed_nodate} suggested a method in the general guidelines for number input and output blocking errors based on three original principles defined by the Institute for Safe Medical Practices(ISMP).The three principles adopted are: \cite{noauthor_chimed_nodate}: 
\begin{itemize}
\item “Do not use trailing zeros for doses expressed in whole numbers” (use 1 mg, do not use 1.0 mg)".
\item “Use zero before a decimal point when the dose is less than a whole unit” (use 0.5 mg, do not use .5 mg)".
\item “Use commas for dosing units at or above 1,000, or use words such as 100 'thousand' or 1 'million' to improve readability”.
\end{itemize} 

The existing T34 user interface applies the first two principles for the output of the  \textit{infusion rate} value. Table 2.1 applies the first principle with 5ml instead of 5.0ml. Nevertheless, in table 2.2 the \textit{rate} value is checked against the second principle with a leading zero before the period. The third principle is irrelevant because the device does not handle numbers superior to a thousand. Such guidelines help to avoid confusion.
 \begin{figure}[h]
\centering
% \begin{table}
% \centering
 \begin{tabular}{||l c||} 
  \hline
 \hline 
 Occlusion & 720mmHg \\ 
 Max. Rate & 5ml/h   \\ 
 Program Lock & ON   \\
 Battery Status & 99\%   \\ [1ex] 
 \hline
  \hline 
 \end{tabular}
 
 \caption{T34 Info Display. No zero on whole number}
% \end{table}
\end{figure}
%  \quad

 \begin{figure}[h]
\centering
% \begin{table}[h!]
% \centering
 \begin{tabular}{||l c||} 
 \hline

 \hline  \hline 
 Volume & 720mmHg \\ 
 Duration & 24:00  \\
 Rate & 0.64ml/h   \\  
 Confirm,  & Press YES   \\ [1ex] 
 \hline  \hline 
 \end{tabular}
 \caption{T34 Info Display. Preceding zero for decimal}
% \end{table}
\end{figure}

Despite these design considerations, navigating the message and instructions on the device might be confusing. Some of the instructions displayed on the screen may mislead a user. An example is confusion between the message and instructions displayed to the user after inserting a syringe into the flange, as shown in Figure 2.4 . The message "\textit{Loaded Correctly}"  indicates the successful insertion of the syringe by the device's flange and plunger sensors. On closer examination of the first line, the message "\textit{Check Syringe}" may be mistaken for instruction by an inexperienced user to check the newly inserted syringe, which is false because it is already loaded. These issues of human factors reinforce the need to ensure not only that the device behaves correctly but also that the user interacts appropriately to enhance the level of post-market safety. 

 \begin{figure}[h]
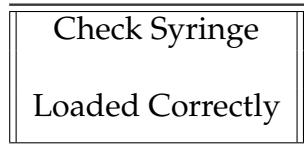

\centering
% \begin{table}[h!]
% \centering
 \begin{tabular}{||c||} 
 \hline

 \hline  \hline 
 Check Syringe \\  
  \\ 
 Loaded Correctly \\ [1ex] 
 \hline  \hline 
 \end{tabular}
 \caption{T34 Screen Display. A confusing message and instruction}
% \end{table}
\end{figure}

By combining the system behavioural model consistency approach of Harold et al.  \cite{masci_formal_2014} and the User Interface of  \cite{harrison_verification_2017}, we can create a refinement strategy to deal with both the system and display states of the T34 \cite{ruksenas_developing_2014}. This refinement method will facilitate the segmentation and linking of the usability requirement of the controller with the appropriate behavioural model of the device user interface. It will facilitate implementation in verification systems such as Spark, which supports subprograms or packages.

%-----------------------------------
%	SUBSECTION 2
%-----------------------------------

% \subsection{Subsection 2}
 
%----------------------------------------------------------------------------------------
%	SECTION 2
%----------------------------------------------------------------------------------------

\section{SPARK Ada}

\subsection{Overview}

 Ada is a strong super type language widely used in critical systems given the strong rule enforcement that prevents many common programming language design mistakes and the introduction of errors such as runtime or out-of-bounds. A potential out-of-bounds error would be the T34 validating a volume of medication greater than the maximum of 22ml allowed for the 30ml syringe type \cite{noauthor_t34_nodate}.  This type of error would eventually result in a thrown exception. Such attributes match the critical systems' requirements where no undefined behaviour should pass safety validation \cite{adacore_spark_guidance_12_webpdf_nodate}. The Ada programming design generally covers three main principles: reliability, efficiency and maintainability  \cite{mccormick_chapin_2015}.  These characteristics of the Ada language make it ideal for static analysers. A program may use  A static analyser to detect the violation of its properties . The static analyser then guarantees that these properties are not violated. The former is called a bug finder, and the latter a \textit{Verifier}. In the case of verifiers, although some initial work is required to help guide the verifier's provers, they report fewer false alarms, which increases confidence in the software. 
 
 SPARK   as a \textit{Verifier} has proven its reliability in many safety-critical system domains, from avionics to space to medical, as a subset of Ada. Spark's restriction introduces accurate static analyses of the code representing both the requirements and the specification of its subprogram unit through a formal logic set \cite{harrison_verification_2017}. In SPARK Ada, the contract expresses such a representation by binding each package or sub-program using appropriate pre-conditions and post-conditions. These conditions ensure that the program will always conform to the given specification. Compared to other verification systems, Spark Ada has demystified the industry's fear by abstracting the formal verification process through automatic provers and providing many integrated tools in one place. To fully understand the implementation of the T34 design pattern, this chapter will discuss the structure and key features offered by the SPARK verifier.

\subsection{ Adacore's spark framework}
 \label{Adacore's spark framework}
 Adacore and Altran created the SPARK framework, which is maintained on an ongoing basis by Adacore. SPARK provides a more secure subset of Ada and static data flow error detection  \cite{mccormick_chapin_2015} . It, therefore, makes verification of difficult or costly program properties tedious compared to methods such as testing or code reviews. The safety offered by the verifier came from the guarantee that, no matter how complex the requirements, most properties and conditions will always hold and be verified \cite{adacore_spark_guidance_12_webpdf_nodate}. As mentioned by Chaplin and Cormick, the process of \textit{formal verification }in SPARK is to evaluate a program without executing the code, hence the name \textit{static verification} \cite{mccormick_chapin_2015}. This process is deterministic due to the underlying mathematical analysis of the program instructions. These guarantees are classified into the following five successive levels of analysis  \cite{adacore_spark_guidance_12_webpdf_nodate} :

\begin{itemize}
\item \textit{\textbf{Stone}} - Verified by accepting the programme code as conforming to SPARK Ada subset.
\item \textit{\textbf{Bronze}}: This stage focuses on the absence of errors in the flow analysis and covers a large portion of the code.
\item \textit{\textbf{Silver}} - Successfully proves the absence of run-time errors (AoRTE) during program execution, such as the absence of out-of-bounds errors.  
It is the typical level for satisfying safety-critical systems, but it varies greatly depending on the available cost or limitations of the systems. 
\item \textit{\textbf{Gold}}  -  Successful verification of programme specifications and its defined contracts. Usually, only a  part of the source code achieves this verification level due to the security level of specific program properties.
\item \textit{\textbf{Platinum}} - As the highest level, this is complete proof of the entire program.
\end{itemize}

  Incrementally approaching these levels allows the developer to easily define and update the necessary primary contracts as the project evolves. In general, a SPARK program is divided into packages containing subprograms. A subprogram does not differ from the usual name for methods in other programming languages, and a subprogram is any instruction with proper executable instructions. There are two types of subprograms in SPARK; the first is the \textit{Procedure}, which does not return any value, while the second type, the \textit{Function}, returns a typed value.  The contracts for subprogram verification such as Depends, Pre, Post and Global can be added to the subprogram specification file or attached to its body, as shown in Figure 2.5 from Mccormick and Chapin's book \cite{mccormick_chapin_2015}. On line 1, the pragma annotation aspect tells the prover that GNATprove should analyse the package. On line 5, the "Selection$\_$Sort" subprogram procedure is declared. From lines 8 to 10, the contracts help the prover generate the proofs.

 \begin{figure}[h]
 \begin{lstlisting}[language= Ada]
 pragma Spark Mode (On);
 ...
 type Array Type is array ( Positive range <>) of Integer;
 ...
 procedure Selection_Sort (Values : in out Array Type)
  --Sorts the elements in the array Values in ascending order
 with Depends => (Values => Values),
      Pre => Values'Length >= 1 and then
                Values' Last <= Positive'Last ,
      Post => (for all J in Values'First .. Values'Last - 1 =>
                Values (J) <= Values (J + 1)) and then
                Perm (Values'Old, Values );
 \end{lstlisting}
 \caption{Example of SPARK Procedure with defined contracts}
 \end{figure}

 \subsection{  Proofs and Flow analysis} 
 \label{section:Proofs and Flow Analysis}

 As mentioned in the previous section, SPARK has removed and added features to Ada for robust program code analysis. Two features cover this analysis; the first is Flow Analysis, and the second is proof. SPARK's GNATprove tool supports both of these analyses. Flow analysis allows data and flow dependencies to be specified in the subprogram. The data dependency usually expresses a relationship between some global data and indicates whether the subprogram should read or write to it. In Figure 2.5, the variable "Values" defined as "in out" indicates that it is both input and output data; therefore, the procedure can overwrite it. The "Depends" contract defines the flow dependency indicating that the procedure data "Values" depends on the current value of "Values". The execution of the GNATprove flow analysis will evaluate the veracity of these contracts.
 
 The Proof Mode of GNATprove provides access to static analysis where the prover asserts all contracts and annotations are transformed into logical statements. Mccormick and Chapin indicate that these statements form the conjectures commonly referred to as verification conditions \cite{mccormick_chapin_2015}. The prover, therefore, assumes that these conditions are  "True but not yet proven". Adopting this mindset at the beginning of the software development life cycle is safe to assume that the test phase can be optimised or even reduced. Further research by Amey \cite{amey_correctness_2002} argues that in the early 20s, adopting the Spark for correcting critical systems can reduce costs and testing by 80 per cent. Analysing 20 years of theorem proving evolution with SPARK,  \cite{chapman_are_2014}, Chapman and Schanda argue that there has been an increase in the use of formal method analysis in the aerospace industry. They also proved that they achieved the expected verification objective on several projects. They significantly reduce additional verification activities such as testing or reviews.

 \section{Summary}

To achieve such verification for T34, the various user interface verification models explored earlier in this literature chapter will first help encompass the system's entire behaviour from the perspective of a working prototype such as the PVS prototype in Pvsioweb. Capturing the user interface (UI) at each state or behaviour of the device system will help to prove that the UI conforms to the requirements and specifications. Using the refinement approach, the derived model will help simplify the system and UI states against the safety and hazard guidelines previously established for the Generic Infusion Pump model.

Implementing a refined generic infusion pump model for the T34 syringe pump in SPARK Ada will contribute to the knowledge of the reference implementation body for the Generic Infusion Pump project. More importantly, it will allow the author to simulate and verify the T34 user interface design model against an existing standard safety risk testbed for GIIP, potentially expanding the scope of the design error assessment. 
% Chapter Template

\chapter{Methods and Implementation} % Main chapter title

\label{Chapter3} % Change X to a consecutive number; for referencing this chapter elsewhere, use \ref{ChapterX}

%----------------------------------------------------------------------------------------
%	SECTION 1
%----------------------------------------------------------------------------------------

\section{Overview }
\label{Overview2}
This chapter discusses the implementation of this project and provides data for some initial qualitative and quantitative analysis.  The author takes an experimental approach by examining the feasibility of implementing the T34 system in SPARK Ada.  The case study methodology is adopted for this experiment.  The development of the device model will form the basis for the analysis of the theoretical aspect of the behavioural model.  A simplified version of the usability requirements of the Generic Infusion Pump is created.  One of the advantages of the Generic Infusion Pump initiative is that it provides a collection of the latest resources needed for an infusion pump.  Since this research is based on a safety-critical device, a crucial step at the beginning of this project is risk analysis.  Fortunately, a generic hazard analysis study captures most of the latest safety concerns of researchers, companies and patients: \cite{zhang_hazard_2010}.  These simplified usability requirements will be merged with the T34 syringe manufacturer's user interface model based on refinement techniques.  As expected, these physical requirements of the T34 syringe are absent for proprietary reasons, and a reverse engineering refinement technique described previously in the background work will be applied.  The newly segmented user interface will represent the different packages or subprograms required to implement a prototype in SPARK Ada.  

 The incremental approach to the development of SPARK Ada has justified the choice of an iterative software development method such as Agile.  Agile is a prevalent development method that reduces costs and facilitates updates to existing requirements as the project evolves\cite{leau_software_nodate}. As this project is intended to serve as a potential case study for the implementation of SPARK in business, the Agile Software Development Lifecycle allows for replicating some of the industry strategies considered overhead.  This context leaves space for testing whether modifying the SPARK contracts for the different subprograms and packages of the refined model will improve the time cost in testing.

  Using SPARK Ada, the generated verification conditions can further support the validation and requirement of the model against the actual device and provide the quantitative results needed to conclude the effectiveness of our model implementation.  Tracking all faults found during implementation and specification will expose the potential severity of faults through analysis.  Another source of data collection is the resources available on websites such as Generic Infusion Pump Project and Adacore\cite{noauthor_generic_nodate} . Analysis of the reported faults and evidence found in any subprogram or package iteration will help assess whether Spark Ada is suitable for such verification work.  This assessment will determine whether an additional iteration is required to achieve the intended goal.  A given UI model subprogram will be considered proven if it has achieved the highest possible level of proof in Spark Ada and the corresponding level of assurance, as discussed in Section \ref{Adacore's spark framework} . An anticipated risk is the strict limitation of the Spark Ada subset for compliance with certain user contracts.  The specific function or behaviour will be shipped as a separate rewritten package to mitigate this risk and ensure that the contract or requirements are met. 
  
  \subsubsection{Ethical Considerations}
As the project resources include access to NHS and UK healthcare resources, any actual future development for training purposes using the final project must be ethically reviewed and approved by the Swansea University COS Ethics Committee to ensure appropriate safety.

\subsubsection{Timeline}

The timeline for this project is divided into three main phases.  The initial phase focused on establishing the background work and research question.  During this stage, it established the initial scope and the underlying motivation of the research project.  The second phase explored the detailed aspects of the structure and methodology of the research, as well as experimenting and finding the right tools for implementation.  This phase was the most time-consuming, given the need to obtain the specification and annotations for the automatic proofs.  Furthermore, in the last phase, the SPARK proofs are evaluated and improved to progressively match the verification objective of the proposed system model for T34.

%----------------------------------------------------------------------------------------
%	SECTION 2
%----------------------------------------------------------------------------------------

\section{Case study : a T34 syringe control model }

\subsection{System Architecture}
\label{Ch3:System Architecture}
The GIP research project provides a fundamental system architecture compatible with legacy and existing market devices  \cite{zhang_hazard_2010}. Such an architecture is generic and component-based, which means that the implementation of any actual device is possible by simplifying the design.  For example, an insulin pump may differ from a subcutaneous analgesia pump.  However, both incorporate some form of an infusion for drug delivery to the patient.  An internal mechanism of the delivery system controls the infusion.  This level of design abstraction ensures that all the safety issues already addressed in the generic are also applicable to the refined system model of the device. In figure 3.1 the proposed architecture outlines a simplified architecture based on the components of the generic insulin infusion pump (GIIP) and the generic patient-controlled analgesic (GPCA).  In the project context, the system architecture only covers the primary control of syringe delivery.  It justifies the purpose of the T34, which is primarily an ambulatory syringe driver with limited key entry and no patient data collection.  The dashed connector line in the diagram represents the data exchange flow through the system, and the thick line represents the physical connection.  Four main components, based on the physical characteristics of the T34, are essential for successful drug delivery: the syringe, the actuator and system control unit, the user interface and the local syringe database.  The T34 only supports a specific range of syringe types, so the local database stores the syringe profiles.  The system controller can access this data to verify the control and command of the actuator.  The actuator monitors the syringe in the syringe unit via the available sensors.  This detection is possible using a set of 3 physical sensors, as shown in Figure 2.1, which track the appropriate type of syringe inserted into the device. 
 
\begin{figure}[h!]
% \centering
\includegraphics[scale=0.22]{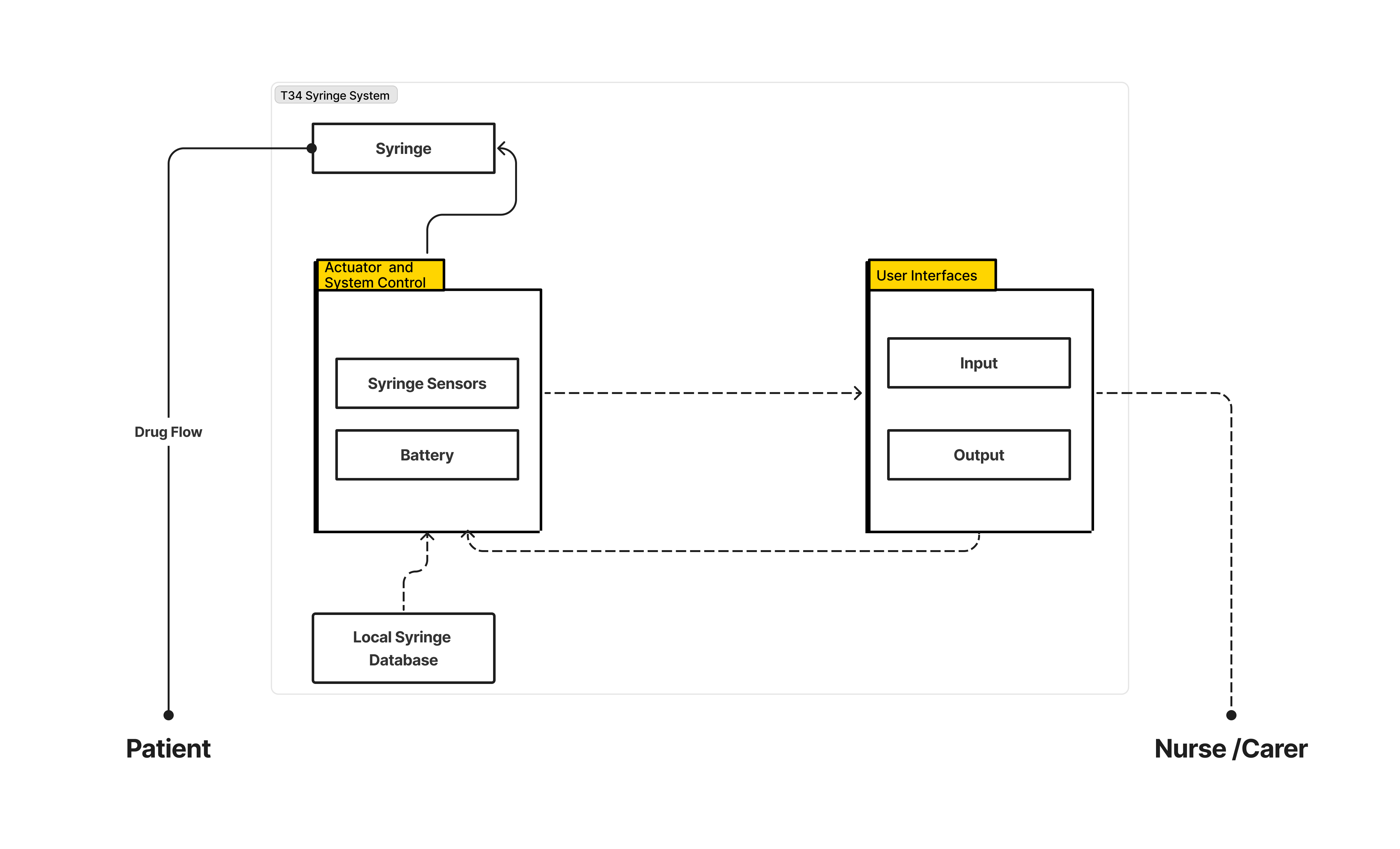}
\decoRule
\caption{ A simplified architecture for T34 based on the GIP system architecture  }
\label{fig:T34 System Architecture}
\end{figure}

Unlike the GIP, the nurse or practitioner does not have to enter a prescription or drug setting.  The input to the User Interface only validates messages or controls information about the device.  Please note that in the system architecture, the term " \textit{User Interface} " refers to all the input or output interactions of the user.  The output sub-component of the user interface represents the on-screen display of the message or instructions and the status of the light, as shown in Figure 2.1. The notification interfaces do not handle the notification alarms to simplify the system.  Most components of the environment are not relevant to the physical characteristics of the T34.  Therefore, external elements such as temperature, pressure and sound are not considered in the scope of the system.  The system controller is the system's heart as it interconnects the different components and handles the necessary logic.  It tracks and processes all input commands from the user and ensures the appropriate result action.  It maintains the syringe presets and all validation parameters for each type of authorised syringe.  Depending on the system input, it sends appropriate instructions or alerts to the user via the display or light status. 

One of the main safety aspects of this architecture is that the user does not enter any drugs or prescriptions into the device.  Therefore, the syringe type recommendation helps the system controller to determine the appropriate automatic infusion rate.  Any error here will impact the patient and violate the device's safety.  Hence the need also to adapt the existing Generic Hazard analysis\cite{zhang_hazard_2010} for the GIP to our simplified architecture.
\subsection{Hazard and Risk Analysis}
In general, the design of infusion pumps has highlighted the complexity of the research for developing medical devices and requires challenging ongoing regulation.  The Generic Hazards provided in the GIP are collected from multiple input sources such as device manufacturers, pump users and clinicians, ISO standard event reports and the Diabetes Technology Society workshop \cite{zhang_hazard_2010}. Much of the insulin pump hazard is based on the ISO14971 guidelines \cite{noauthor_iso14971_nodate}initially established in 2007 but updated in 2019. The \textit{risk} associated with infusion pumps is presented in the ISO as the result of the probability of occurrence of the \textit{harm} and its severity covered by the defined risk acceptability criteria. The impact of the consequences of a risk-associated hazard defines its severity.  As a critical device, any resulting injury or damage to people's health is considered harmful to the device. To properly understand and address a hazard, it is essential to identify all potential sources of injury or damage within the device or its environment.  As detailed in the Appendix   \ref{fig:hazardoussituations}, hazards derive from a sequence of events generated by a hazardous situation which consequently has a potential for harm.  There are five main categories of hazardous events that correspond to most medical devices \cite{zhang_hazard_2010,noauthor_iso14971_nodate}:
\begin{enumerate}
\item Therapeutic
\item Energetic
\item Chemical or Biological
\item Mechanical
\item Environmental
\end{enumerate}

Hazards for any risk can arise from any part of the entire development of a device, including its development, installation, maintenance and disposal.  For example, the T34 is set up for 24-hour administration; consider a scenario where the patient needs to take a shower.  The device must have measures to reduce the risk of reverse flow due to syringe connector siphoning. The \textit{Hazards Analysis} document documents some of the significant hazards of the GIIP model and therefore reflects the general mistakes of small portable devices\cite{zhang_hazard_2010}. This document is a basis for the safety reference standard for all future designs.  Therefore, basing our implementation on this document, we reflect these considerations for the T34 syringe pump and potentially address future device upgrades.

In the following table (\ref{table:T34hazardsandrisk})  the author presents a non-exhaustive list of the main hazards applicable to the system architecture proposed in Section \ref{Ch3:System Architecture}. In the model's scope, only hazardous situations' operational, software and hardware sources are considered to determine the hazards.  The hazardous situations covered mainly to expose the user interfaces via their input and output units.  In many cases, most foreseeable event sequences can lead to several hazardous situations simultaneously, so a single hazard is included for simplicity.

\begin{table}[h] 

  \centering
  \begin{tabular}{p{3cm}llp{3cm}}
    \toprule
    \textbf{Hazard}&\textbf{Sequence of Events}&\textbf{Hazardous Situation}&\textbf{Harm}\\
    \midrule 
	Fluid Pressure & 		Air in line	 & Overdose & Injury \\
    \hline
    Fluid Pressure & 		Low resistance flow & Underdose & Unintended medication  \\
      \hline
 Syringe Data Availabitlity	&Error reading data	 & Underdose  &Unintended medication \\
   \hline
  Syringe Data Integrity	& Invalid Syringe Type 	 & Underdose  &Unintended medication \\  \hline
  Voltage	& 	Battery draining  Fast		 & Underdose & Injury \\
    \hline
  Voltage	& 	User ignored alarm		 & Device turned off & Unintended medication \\
    \hline
  Alert & Led lights  inconsistently & Nuisance  alarms &Unintended medication \\
    \hline
  Alert &Confusing Message & Nuisance  alarms & Unintended medication \\
    \hline
  Critical Perfomance&  Overflow error  & Syringe Blocked& Injury \\
    \hline
  Delivery quantity&Incorrect prescription& Insuficient Drug volume & Serious injury \\
  & & & \\
  & & & \\
  \bottomrule
\end{tabular}
\caption{T34 Refined Risk and Hazards analysis }
  \label{table:T34hazardsandrisk}
\end{table} 
As with any risk, appropriate control measures should be implemented to mitigate or reduce the risk to an acceptable level where it is a residual risk.  These measures can be implemented in several ways.  For mitigation, the verification of the implementation and its effectiveness is possible from the beginning of the development.  The correctness of the SPARK contract asserts this verification.  It again supports the underlying project objective of proving that formal methods are effective with less overhead for commercial adoption.  Another control measure, as with any system, is to ensure that guidelines and support manuals cover residual risks to enable staff to deal with any unexpected risks.
 
\subsection{Safety Requirements}
 The hazard outlined in the previous section provides an overview of the operational conditions and some key behaviours that may occur for the T34 syringe driver.  A system requirement, in general, covers the behaviours, functionality and capabilities of the system  \cite{palumbo_requirements_nodate,SCHMIDT2013113}.  It is essential to consider these requirements from the user's point of view and their expectations regarding the challenges they face.  Such an argument is crucial when considering the level of stress on medical staff as discussed in Section \ref{Device Characterisitics}.More importantly, it defines the potential constraints that can bind and limit the system.  Modelling these behaviours allows the creation of specific and measurable tasks that represent what the device should do and how it should perform it  \cite{SCHMIDT2013173,SCHMIDT2013185,palumbo_requirements_nodate}.  In his analysis of requirements management for a safety-critical system, Palumbo \cite{palumbo_requirements_nodate} presented the following eight rules for characterising a reasonable system requirement:
\begin{enumerate}
\item \textit{Necessary} -  The requirement states an essential capability or physical characteristic 
\item \textit{Concise} - The requirement is simple and contains no justification or description of the use of the system.
\item \textit{Complete(Standalone)} - No further implications, a self-sufficient requirement.
\item \textit{Consistent} - No contradiction or duplication of another requirement.
\item \textit{Unequivocal} - No confusion, only a single interpretation for each requirement. 
\item \textit{Feasible} - Implementation of the requirement is possible despite limitations.
\item \textit{Verifiable/Testable}  
\item \textit{Traceable} - The requirement is uniquely  identified  and  coded
\end{enumerate}

 The proposed T34 architecture is then evaluated to ensure safety while ensuring compliance and consistency with the defined requirements.  These requirements cover all structural aspects of the software, hardware, system and user interfaces, and they must protect against errors and recover from possible failures.  In addition, they must cover the reliability and availability of the device and identify all states and conditions that must not occur.  Although both software and hardware components can contribute to the safe and healthy operation of the pump, there are cases where they may be ineffective or even unable to mitigate certain risks \cite{zhang_generic_2011}. For example, many environmental hazards for medical devices such as biological are entirely independent of the software.  In such cases, this residual risk must be mitigated by control measures such as training and policies.   
 
As part of the development of the GIIP testbed, the defined risk control measures led to the categorisation of safety requirements for the model  \cite{zhang_generic_2011}. The categories allowed for a clear definition of the infusion administration and removed ambiguities related to scheduling and planning.  They also provide a diagnosis by ensuring a safe event record, which is essential in case of a hazard.  While not requiring results verification, this category is essential for monitoring the reproduction of the hazard and correcting the critical potential.  Finally, the requirements cover most of the hazardous situations previously foreseen for the infusion pump, as covered by the GIIP Hazard Analysis.  In addressing the safety aspect of the requirements, it is easier to bind and adapt the proposed architecture to generic safety requirements.  Therefore, the non-exhaustive list of safety requirements below is derived from the T34 training and safety guidelines defined by  NHS for Education Scotland\cite{nhs_education_for_scotland_guidelines_nodate,noauthor_t34_nodate} and the generic safety requirements of the GIP research project \cite{zhang_generic_2011}. For the scope of the proposed T34 system architecture, only the requirements for infusion administration, user interfaces and alerts are covered.
 
 %%%%%%%%%%%%%%%%%%%%%%%%
% start table 
%%%%%%%%%%%%%%%%%%%%%%%%%%
 \begingroup % Begin a TeX group in order to localize scope of next three instructions
\setlength\extrarowheight{2pt}
\small
  \begin{longtable}{p{2cm}|p{8cm}|p{2cm}}
   
  \hline
  \textbf{Req ID}&\textbf{Software Requirement}&\textbf{Hazardous     Situation}\\
 \hline
\endfirsthead 

\multicolumn{3}{@{}l}{\tablename~\thetable, continued.} \\[0.5ex]
\hline
  \textbf{Req ID}&\textbf{Software Requirement}&\textbf{Hazardous   Situation}\\ \hline
\endhead 

\multicolumn{3}{r@{}}{\em Continued on following page}\\
\endfoot

\endlastfoot
 
 \multicolumn{3}{l}{1.1 Infusion Control} \\ 
   \midrule 
     1.1.1& The flow rate shall remain accurate to the automatically calculated flow rate over 24 hours &  \\
        1.1.2&The pump shall not allow the syringe type to be changed after confirmation unless a new syringe is inserted.  Remind the user to validate any new syringe type inserted and its default settings against the prescription.&  \\
             1.1.3&The pump flow rate for the set syringe shall not exceed (overflow) the maximum flow rate of the device that meets requirement  1.2.5 &  \\
             1.1.4  & The drug delivery shall be evenly distributed over the 24 hours to meet requirement 1.2.1 & \\
      \midrule  
 \multicolumn{3}{l}{1.2  Drug  Setting and Administration} \\ 
   \midrule 
  1.2.1&The duration of the infusion should cover 24 hours of a day&  \\
    1.2.2&The pump shall inform when the recognised syringe type data is set and administer the medicine following the set data profile after checking &  \\
      1.2.3&The unit of volume of the pump must be ml, and the infusion duration must be expressed in hours&  \\
        1.2.4& If the infusion in progress has been interrupted for more than 5 minutes, the pump must trigger an alert each 1-minute up to 1-hour &  \\
          1.2.5&The maximum pump flow rate for all syringe types is 5 millilitres per hour (5ml/hr) &  \\
      
  \midrule  
 \multicolumn{3}{l}{1.3   Syringe Validation - according to the tolerance defined in the  ISO 7886-1:2017\cite{noauthor_iso_nodate} \footnotemark{}\footnotetext{The tolerance for the graduated syringe for this research is as follows: 
\begin{itemize}
\item  5 \% of the Volume of drug expelled for a value greater than or equal to half the nominal capacity (With a nominal capacity < 5 ml). \\
\item   (1.5 \% of Volume + 2 \% of Volume expelled)   for less than half the nominal capacity (With the  nominal capacity < 5 ml).
\item  4 \% for a value less than half the nominal capacity (With the  nominal capacity >= 5 ml).
% \begin{equation}
%     \geq  \text{5 ml}
% \end{equation} ).
\item (1.5 \% of Volume + 1 \% of Volume expelled)  for a value less than half the nominal capacity (With the  nominal capacity < 5 ml) .
\end{itemize}
 } } \\ 
   \midrule 
 1.3.1 & The remaining volume calculation shall be accurate to the calculated tolerance  &  \\
   1.3.2& The remaining volume calculation shall be updated after the primed syringe, considering the average expelled volume for priming &  \\
      1.3.2& If the syringe volume calculation is at 0ml within the tolerance value, the pump shall emit an alert to the user for complete infusion delivery &  \\
  \midrule  
 \multicolumn{3}{l}{1.4  Data Integrity and Availability} \\ 
   \midrule 
  1.4.1&The selection of the syringe profile shall not be saved until the user has reviewed and confirmed its values & \\
    1.4.2&The pump shall initialise and validate critical data like syringe profile for the different supported brands, maximum flow rate and sensor data. & \\
  \bottomrule
  \caption{Requirements on Drug  Administration }
  \label{table:T34safetyrequirementsDrugAdministration}
\end{longtable}

\endgroup % End of TeX group 
%%%%%%%%%%%%%%%%%%%%%%%%
% end table 
%%%%%%%%%%%%%%%%%%%%%%%%%%

 %%%%%%%%%%%%%%%%%%%%%%%%
% start table 
%%%%%%%%%%%%%%%%%%%%%%%%%%
 \begingroup % Begin a TeX group in order to localize scope of next three instructions
\setlength\extrarowheight{2pt}
\small
  \begin{longtable}{p{2cm}|p{8cm}|p{2cm}}
   
  \hline
  \textbf{Req ID}&\textbf{Software Requirement}&\textbf{Hazardous     Situation}\\
 \hline
\endfirsthead 

\multicolumn{3}{@{}l}{\tablename~\thetable, continued.} \\[0.5ex]
\hline
  \textbf{Req ID}&\textbf{Software Requirement}&\textbf{Hazardous   Situation}\\ \hline
\endhead 

\multicolumn{3}{r@{}}{\em Continued on following page}\\
\endfoot

\endlastfoot
\multicolumn{3}{l}{2.1  Resistance to accidents} \\ 
   \midrule 
   2.1.1 & The pump shall have a locking mechanism to protect the syringes.  Only authorised personnel should be able to unlock it.  &  \\
    2.1.2 & The pump shall require a confirmation from the user to change the selected syringe profile when it is a syringe profile. &  \\ 
     \midrule  
    \multicolumn{3}{l}{2.2  User Input} \\ 
   \midrule  
    2.1.2 & The pump shall issue a warning to the user if no input has been loaded within one minute when it is in a state that requires input, for example, setting the syringe type  &  \\ 
     \midrule  
    \multicolumn{3}{l}{2.3  Keypad } \\ 
   \midrule 
   2.3.1 &  The pump shall allow the keypad to be locked and unlocked  &  \\
    2.3.2 & The user shall require a confirmation from the user to change the selected syringe profile.  &  \\
      2.3.3 & The pump shall issue an alert if a key is held down for at least 3 minutes   &  \\
    & &  \\
         \midrule  
    \multicolumn{3}{l}{2.3  Information Display } \\ 
   \midrule 
   2.1.1 & The pump shall display concise and clear information to assist the user during operation.  This information may include the current syringe profile, current flow rate, time remaining to administer the drug, other critical sources of data and warnings &  \\  \\
  \bottomrule 
\caption{Requirements on User Interfaces }
  \label{table:T34safetyrequirementsUserInterfaces}
  \end{longtable}
\endgroup % End of TeX group 
%%%%%%%%%%%%%%%%%%%%%%%%
% end table 
%%%%%%%%%%%%%%%%%%%%%%%%%%

 %%%%%%%%%%%%%%%%%%%%%%%%
% start table 
%%%%%%%%%%%%%%%%%%%%%%%%%%
 \begingroup % Begin a TeX group in order to localize scope of next three instructions
\setlength\extrarowheight{2pt}
\small
  \begin{longtable}{p{2cm}|p{8cm}|p{2cm}}
   
  \hline
  \textbf{Req ID}&\textbf{Software Requirement}&\textbf{Hazardous     Situation}\\
 \hline
\endfirsthead 

\multicolumn{3}{@{}l}{\tablename~\thetable, continued.} \\[0.5ex]
\hline
  \textbf{Req ID}&\textbf{Software Requirement}&\textbf{Hazardous   Situation}\\ \hline
\endhead 

\multicolumn{3}{r@{}}{\em Continued on following page}\\
\endfoot

\endlastfoot
\multicolumn{3}{l}{3.1  Power-on self-test (POST)} \\ 
   \midrule 
 3.1.1& When powered up, the pump shall run a POST to check the equipment, such as the sensors and battery, the LCD screen display, and the LED&  \\ 
  &  &  \\
    &  & \\
 \midrule
 \multicolumn{3}{l}{3.2 Alert and Warnings} \\ 
   \midrule 
 3.2.1& The alert condition and warning shall be visually accessible&  \\
  3.2.2&Any alert or error shall identify the conditions violated which caused the alert.&  \\
   3.2.3& The pump shall be designed to contain a fail-safe mechanism that stops the infusion in any event in event of a critical error&  \\
   & & \\
    & &   \\
  \bottomrule 
\caption{Requirements on Alert, Alarm}
  \label{table:T34safetyrequirementsAlertAlarm}
\end{longtable}

\endgroup % End of TeX group 
%%%%%%%%%%%%%%%%%%%%%%%%
% end table 
%%%%%%%%%%%%%%%%%%%%%%%%%%

In the typical model-based development approach, developers typically use security requirements in two major stages of development. \cite{zhang_generic_2011}. The first stage is Design Verification, in which the design model is implemented and its behaviours checked against the defined requirements for violations.  This strategy identifies defects early before converting the final model into its corresponding software implementation.  This step will be developed in the next Section \ref{model design and refinement} on Model design and refinement. The second step consists of verifying the implementation of the design.  It assesses whether the implementation does not deviate from the initially defined security requirements and does not meet the expected behaviours.  It will be addressed in the implementation in SPARK.  In addition, the usual development process, such as the test suite design, can reinforce the risk by asserting any violation of the expected system outcome.
\subsection{ Model design and Refinement}
\label{model design and refinement}

 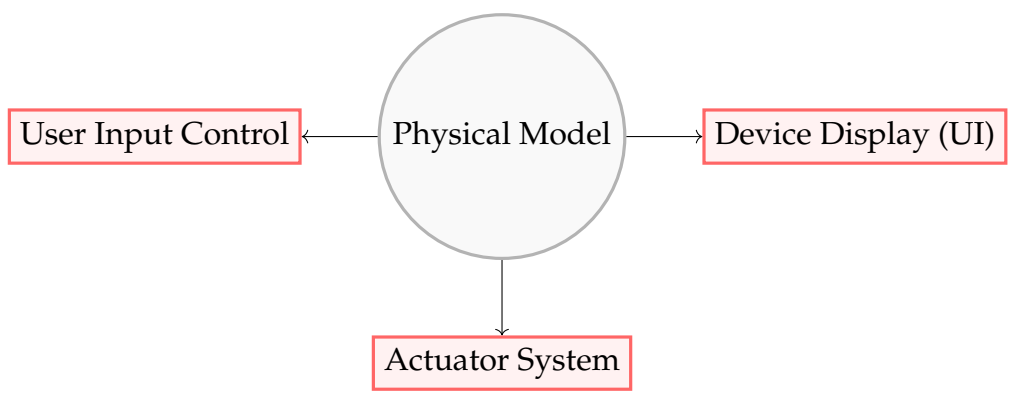
\begin{figure}[h!]
\centering
\begin{tikzpicture}[
roundnode/.style={circle, draw=gray!60, fill=gray!5, very thick, minimum size=7mm},
squarednode/.style={rectangle, draw=red!60, fill=red!5, very thick, minimum size=5mm},
]
%Nodes
\node[roundnode]      (maintopic)                              {Physical Model};
\node[squarednode]        (leftsquare)       [left=of maintopic] {User Input Control};
\node[squarednode]      (rightsquare)       [right=of maintopic] {Device Display (UI)};
\node[squarednode]        (lowercircle)       [below=of maintopic] {Actuator System};

%Lines
\draw[->](maintopic.west) --  (leftsquare.east);
\draw[->] (maintopic.east) -- (rightsquare.west);
\draw[->] (maintopic.south) -- (lowercircle.north);
\end{tikzpicture}
 \caption{Refined T34 Model components diagram.}
 \label{fig:t34 model components}
% \end{table}
\end{figure}
 
 The purpose of designing a model is to encompass the behaviour of a system.  These behaviours consist of how the device behaves and under what constraints certain behaviours are expected.  In the first step, the physical model of the T34 syringe pump is designed and represents its abstract representation according to the defined system architecture.  Then, a refined model is proposed based on the physical model and the previously defined safety requirements.  During the refinement process, the consistency of the system, as stated by Harold et al. in their PVS prototype implementation  \cite{masci_formal_2014}  is verified  against the human factors of the architecture.
 
 The physical model diagram in figure \ref{fig:t34 model components} groups the system architecture into 3 main components:  the \textit{User Input Control} , the \textit{Diplay Device(UI)} and the\textit{ Actuator System}. These three core components are sufficient and essential for any infusion pump implementation.  They also match the components previously defined in the GIIP and GPCA model.  they should therefore be sufficient to conclude our implementation in SPARK Ada.  Then facilitates the incremental development of the necessary verification conditions for each core component and limits the existing abstraction by similar features. 
\begin{itemize}
\item The \textit{User Input Control} integrates keypad or button inputs and associated logic behaviours or outputs throughout the device's system execution.
\item The \textit{Device Display} represents the actual screen display and all its parameters.  It is one of the output units of the system architecture, as it provides information, instructions and commands to the user.
\item The \textit{Actuator System} integrates the sensor unit for syringe detection and control, infusion measurement and drug delivery.  It is a crucial component as it presents the most critical hazard risk.  Any failure of the actuator movement is critical to the patient, whereas a failure of the device display is not always dangerous to the patient.  An atypical case could be a warning alert that may not be critical or may be related to an underlying process, such as a logger diagnosis.  If it is ignored or does not appear to the user on the screen, it will not be critical until the pump is reported as part of an incident for which the Medical Physics Department may want to explore the diagnostic data.
\end{itemize}

The previous model mainly covers the inherently physical aspects in its abstraction, so it is essential to have an improved model that adds the constraints that the T34 pump should bind to and its underlying behaviours.  In the prototype GPCA model verified by PVS and its corresponding model design process\cite{masci_formal_2014,kim_safety-assured_2011}  the use of states is explored to express the system at a particular point in its execution.  It is designed through the use of a state controller, which is an implementation of a state flow.  In our case, a typical \textit{State Machine} can effectively represent the system's states and behaviours, including the user interface's state and the corresponding inputs at each state. A \textit{State Machine } is an abstraction with a finite or exact number of states that represent the underlying behaviour of a system at a given time.  Graphically, Harel's Statechart\cite{harel_david_statechartspdf_nodate}widely used in the Unified Modelling Language (UML), provides an unambiguous representation of a state machine.

The state diagram in the figure \ref{fig:state diagram} contains an initial state {$s_0$} and a list of intermediate states identified by {$s_i$}. Each state is connected to at least one other intermediate state by a \textit{transition}  indicated by the direction of the arrow. A \textit{transition} is 
  a behaviour which triggers a change of state.  Each transition is the product of an input trigger ( Keypress) or a condition for specific requirement constraints ( Safety requirements) and the User Interface ( Display).  The following general equation defines the label transition system.
 \begin{equation}\label{eq:labelsystem} 
     \mathscr{A} = 2^E \times C \times 2^A  
  \text{ with } \mathscr{A} => \text{label transition}  
 \end{equation} 
 where:
 \newenvironment{conditions}
  {\par\vspace{\abovedisplayskip}\noindent\begin{tabular}{>{$}l<{$} @{${}={}$} l}}
  {\end{tabular}\par\vspace{\belowdisplayskip}}

\begin{conditions} 
 E     & Input events /Trigger \\
 C    &  Conditions /Constraints \\   
 A  &  Actions/Output
\end{conditions}

Each state receiving the transition in the direction of the arrow is considered the next state for the transition.  The state at the end of the arrow is the previous state for the transition.  Each state coded in the diagram is associated with the behaviour of the corresponding state in human language for better readability.  The label of the transition reads as follows:
"\textit{Input  \textbf{[} Condition1,... Condition n\textbf{]/} Message or Instructions}". For example, the label of the transition between the state \textbf{ $s_5$} to \textbf{$s_6$} reads as follows: \textit{Yes/ "Start Infusion"}. It means that the expected input trigger is the "Yes" button on the device, which will update the display with the instructions  \textit{"Start Infusion?"}.After the transition, the machine's current state is set to \textbf{$s_6$}.The following two transition labels of the state \textbf{$s_6$} are essential to understand why this is a refined model that successfully merges the physical model and the safety requirements. 
\begin{enumerate}
\item From \textbf{$s_6$}  to \textbf{$s_7$},the current state of the UI is the instructions \textit{"Start Infusion?"} therefore, if the user presses the Yes button, the UI is updated to the new message \textit{"Pump  Delivering"}. The UI state change has then been validated and accounted for in the global state machine.
\item From \textbf{$s_6$}  to \textbf{$s_8$}, the UI state is always the instruction \textit{"Start Infusion?"}. Therefore, the instructions have validated the state of the user interface, now the condition  \textit{ [Timeout(2mn)]} guarantees the safety requirements 3.2.1 and 2.1.2 by verifying that the user confirms the previous instructions between the time required for security.  If not, the new state of the User Interface is validated by the action represented by the alert message \textit{"Pump Paused too Long"}.
\end{enumerate}

  \begin{figure}
\centering
  \begin{tikzpicture}[shorten >=1pt,node distance=2cm,on grid,auto]
  \tikzstyle{every state}=[fill={rgb:black,1;white,10}] 
  \node[state,accepting,initial,accepting above, accepting text = {end}] (s_0)  {$s_0$};
    \node[state,below of= s_0] (s_1)  {$s_1$}; 
    \node[state,below of= s_1,yshift=-3cm] (s_2)  {$s_2$};

  \node[state,right of= s_1,xshift=2cm] (s_3)  {$s_3$};
    \node[state,below of= s_3,xshift= -4cm,yshift=-5cm] (s_4)  {$s_4$ }; 

   \node[state,right of= s_4,xshift=1.5cm, yshift=4cm] (s_5)  {$s_5$}; 
    \node[state,right of= s_5, yshift=-3cm] (s_6)  {$s_6$};

    \node[state,right  of= s_6,yshift=2cm, xshift= 2cm] (s_8)  {$s_8$};
      \node[state,above of= s_8,yshift=3cm] (s_7)  {$s_7$};
            \node[state,above of= s_7, xshift=2cm,yshift= 4cm] (s_9)  {$s_9$};
     \node[state,above of= s_3, yshift=4cm] (s_1_0)  {$s_{10}$};        
            
            \draw (s_0)   edge[above] node{} (s_1)
             (s_1)   edge[above ] node{} (s_2)
              (s_2)   edge[bend left,above] node{} (s_1)
              (s_2)   edge[bend left,above] node[sloped,above ]{\small No/Message:{Preloading}} (s_3) 
              (s_3)   edge[ above] node[sloped,above ]{\small [Clamp Down]/"Loaded Correctly"} (s_4)
               (s_3)   edge[ loop left] node[sloped,above ]{\small Back/(plunger pos+ 1)} (s_3)
               (s_3)   edge[ loop right] node{\small [Preloading]/"Load Syringe"} (s_3)
              (s_4)   edge[ above] node[sloped,above ]{\small Yes/"Confirm, Press YES"} (s_5)
              (s_5)   edge[ above] node[sloped,above ]{\small Yes/ "Start Infusion?"} (s_6)
              (s_6)   edge[ above] node[sloped,above ]{\small Yes/"<<Pump delivering"} (s_7)
              (s_6)   edge[ bend right,below] node[sloped,below ]{\small [Timeout(2mn)]/"Pump Paused too Long"} (s_8)
              (s_8)   edge[ bend right,below] node[sloped,above ]{\small Start/"<<Pump Delivering"} (s_7)
              (s_7)   edge[bend left, below] node{} (s_3)
              (s_3)   edge[bend left, below] node[sloped,above ]{\small [Plunger Pos = 0]/"Delivery Done"} (s_9)
              (s_9)   edge[bend left, below] node[sloped,above ]{\small Yes/"Resume Successfully"} (s_7)
              (s_3)   edge[bend left, below] node{} (s_1_0)
              (s_1_0)   edge[bend left, below] node{} (s_3)
              (s_9)   edge[bend right, above] node[sloped,above ]{\small off/} (s_0)
              (s_1_0)   edge[ loop above] node[sloped,above ]{\small "Charge Battery"} (s_1_0)
       ;

%   \path[->]  (s_0) edge  [loop above]  {0,1} ( );
\end{tikzpicture}

\begin{tabular}{||c c ||} 
 \hline
 States & Associated Behaviour Name \\ [0.5ex] 
 \hline\hline
 $s_0$ & OFF \\ 
 \hline
 $s_1$ &IDLE \\ 
 \hline
 $s_2$ & PRELOADING  \\
 \hline
 $s_3$ & ACTUATOR ON   \\
 \hline
 $s_4$ & SYRINGE LOADED   \\
 \hline
 $s_5$ & SYRINGE VERIFIED   \\ 
 \hline
  $s_6$ & SYRINGE CONFIRMED \\ 
 \hline
 $s_7$ &  INFUSION STARTED \\ 
 \hline
 $s_8$ & PUMP PAUSED  \\
 \hline
 $s_9$ & INFUSION STOPPED  \\
 \hline
 $s_{10}$ & PUMP INFO  \\[1ex] 
 \hline  
\end{tabular} 
 \caption{A state diagram of the refined model of UI and system Behaviours.}
 \label{fig:state diagram}
% \end{table}
\end{figure}
  
With the previous refined model, it became apparent that the global design model of the T34 was manually verified.  As expected to complete its second verification step, the model has all the necessary information for complete implementation in SPARK.  The following section explores the implementation of the model in SPARK and any implications in terms of supporting resources.
\subsection{ Resources and Implementation } 
As important as the refined model may be, a successful implementation is highly dependent on available resources and appropriate materials.  Such a decision is made wisely as it also affects the completion and management of the research and the way the author addresses the issues raised.  Implementation is mainly covered in the following two subsections:  \textit{Materials and Setup }and \textit{ SPARK Implementation}. 
\subsubsection{Materials and setup}
\label{materials and setup}
For the development hardware, a Lenovo laptop with the following specifications:
\begin{itemize}
\item \textit{16GB} of RAM memory
\item  An\textit{ Intel® Core™ i7-10510U CPU @ 1.80GHz × 8} processor
\item A graphics card \textit{Mesa Intel® UHD Graphics (CML GT2)}
\item A \textit{64-bit} Linux operating system on \textit{Ubuntu 22.04.1 LTS}
\end{itemize}

Adacore supports the Ada and SPARK community with supported and well-maintained tools for code development.  For learning and support issues, they provide membership access to the GNAT Academy, which has provided the author with free access to Ada language  and SPARK  learning resources \cite{noauthor_adacommercial_nodate}. The GNAT Community Edition 2021 version of the GNAT Studio IDE, formerly GNAT Program Studio (GPS), is used to create the development code for the implementation. The version \textit{20210519-103 }of the binary is installed on the development hardware.  This version of the IDE is free for students, developers and other software enthusiasts.  It offers the same look and feel as traditional IDEs and is almost identical to GNAT Pro.  The only difference is that it lacks Pro tools like Codepeer for code analysis and QGen for automatic code generation .  GNAT Studio IDE provides a built-in compiler called GNAT Compiler that checks the code according to the \textit{Stone} Assurance Level mechanism and generates the appropriate executable code\cite{adacore_spark_guidance_12_webpdf_nodate}.  For formal verification of the T34 model, GNATprove integrated with GNAT Studio is used as described in detail in Section  \ref{section:Proofs and Flow Analysis}.It also incorporates the Why3 system \cite{chapman_are_2014} for automatic execution of multiple theorem provers and provides detailed reports on failed proofs.  For GNAT Studio to work correctly, it is recommended that you have at least 1.5GB of space and a minimum of 3GB of RAM for every million lines of code.

The choice of an Ada-based user interface library supporting Spark Ada was a significant challenge in the decision to implement this project.  The author first took a trial and error approach before conducting a comparative test of three main Graphical User Interface (GUI) libraries:  \textit{GTKAda}, \textit{GNOGA} and \textit{ Ada$\_$GUI}. The choice is made according to the learning curve, considering the project's duration, the native nature of the library in Ada, and the excellent advice of the Ada community\cite{pinho_ada_2020} 
\begin{enumerate}
\item \textit{GTKAda}: This is the official toolkit supported by Adacore.  It has a more significant learning curve, which was not ideal given the scale of the project.  Solid knowledge and understanding of GTK are necessary for any implemented project to succeed.  The installation was tedious because, in a Linux environment, you can have conflicts with different versions of GTK installed for other applications.  GTKAda is not a native Ada library.  Instead, it is a C++ binding library but cross-platform, which is ideal for better device support.  However, Its design is not task-safe.
\item \textit{Gnoga}:This is a critical Ada GUI library and toolkit created by David Bottom \cite{noauthor_gnoga_nodate}. It is designed to be a cross-platform and platform independent toolkit. \textit{GNOGA} can also run locally or remotely via a web server on computers and mobile devices.  It is purely written in Ada and supports the Ada 2012 framework.  One of the main drawbacks of \textit{GNOGA} is that it is resource intensive and consumes many hardware resources, as it adds many expensive callbacks to update the user interface based on events.  For this reason, we did not use this library for our graphical prototype of the T34 syringe driver.
\item  \textit{Ada$\_$GUI} : This is a forked and lightweight version of  \textit{GNOGA} created by Jeffrey R. Carter \cite{carter_ada_gui_2022}.It is a very simplified version of \textit{GNOGA}. Compared to a traditional Graphical UI like \textit{GNOGA}, \textit{Ada$\_$GUI}  does not use callbacks, which eliminates expensive calls and opens access to the control thread.  It uses a task-based approach where the GUI communicates via an event queue.  It is well documented and open-source.  The code is properly annotated, making it easy to understand if extensions are needed.  It runs by default on a web server but inherits all the core \textit{GNOGA} functionality for local deployment on specific hardware, which is ideal for extending this research study from a simulation prototype to a real hardware prototype.  For these reasons, the prototype is implemented with this GUI library.
\end{enumerate}

As with any standard development project, code management is crucial for managing changes and tracking issues throughout the development of a project.  For code management and version control, Git via the GitHub platform is used to host the code repository for versioning code and fixing issues.  An advantage of Git is the fail-safe approach to easily switch to different project states or branches, which is also accessible locally and remotely.  An agile methodology facilitates the iterative development of the prototype, with the author creating the initial packages representing the architecture and progressively integrating the patch tests.  Microsoft Excel is used for project management via the Gantt chart. 

In terms of overall budget and resources, this research did not require the purchase of additional equipment to develop the prototype.  All the necessary software and verification tools are available free of charge via the websites of the relevant suppliers.  In order to avoid the risk of hardware or compatibility problems, the author had access to the Computational Foundry laboratory at Swansea University to use the pre-installed GNAT Studio.  All the required resources were gathered to proceed with the implementation of the T34 model code, as described in the next section.
At the end of the research, a completed and successfully verified prototype user interface in SPARK Ada and a document to help scale the project should be provided.    
\subsubsection{SPARK implementation}
\label{SPARK implementation}
\begin{figure}[h]
\centering
\includegraphics[scale=0.4]{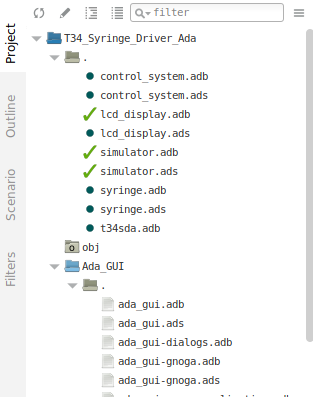}
\decoRule
\caption{Project Code Structure   }
\label{fig:Code structure}
\end{figure}

The code implementation is structured to be easily readable and self-explanatory to match the initial physical model of the system architecture. The figure \ref{fig:Code structure} presents the chosen architecture. The \textit{Ada$\_$GUI} library toolkit is included and loaded as a library to separate its main source code from the actual prototype implementation code.  Some of the library's GNOGA dependencies are also visible.
In the root path of the "T34$\_$Syringe$\_$Driver$\_$Ada"project's main folder, the \textit{Obj} folder contains all the binaries compiled to deploy and run the project via its web server. \textit{Ada$\_$GUI} folder contains three main files at compile time to successfully deploy the Prototype user interface: a \textit{boot.js}, \textit{boot.html } and a\textit{ jquery.min.js}. The other files in the root path are our main SPARK Ada package files for the prototype.  They are named after the model unit or components represented in the main physical model.  As a reminder, a typical SPARK package has \textit{.ads} and \textit{.adb} files representing the specification and body of the package, respectively.  For example, the "\textit{lcd$\_$display}" package, which represents the actual screen of our prototype device, has all its contracts specified in the specification file and its logic managed in the \textit{.adb} body file. The "\textit{syringe}" package manages the local database of available and supported syringe types and their corresponding profiles to be loaded when the pump is running. The "\textit{control$\_$system}" package covers the pump's input and actuation control system.  As the central unit of the package, it made sense to integrate the state controller for the refinement model into this package, as all other packages interact with it.  As the developed prototype simulates a real device, all the simulation generators for a number of the real devices are centralised in one package. The \textit{t34sda.adb} s our main package which integrates and initialises the \textit{Ada$\_$GUI },the control system , the LCD display and simulator.

\begin{figure}[h!]
\centering
\includegraphics[scale=0.4]{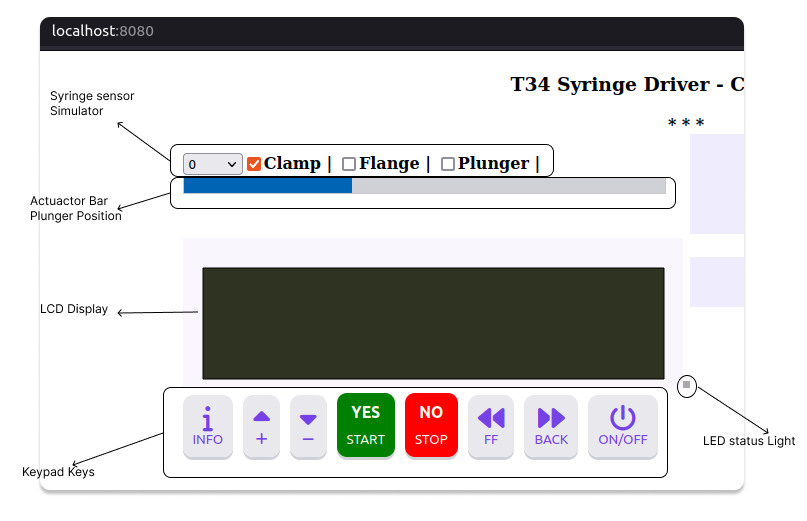}
\decoRule
\caption{Final Prototype in SPARK Ada  }
\label{fig:Final prototype OFF}
\end{figure}

To better understand the code aspect of this study, it is essential to understand the prototype built with the corresponding code.  Below is a figure(\ref{fig:Final prototype OFF}) of the final prototype that is currently deployed and running on the web server.  The prototype is in the $s_0$ state (OFF) representing the model's state according to the running system.  Here, most user interface components are similar to the actual device in size and style, making the prototype familiar to potential future medical users.  One might notice that the status light is turned off as it is supposed to validate the state of the user interface at this particular stage of the refined model.  The LCD screen is also off, as expected.  Above the LCD screen, a progress bar is visible.  The blue bar's end represents the plunger's last position on the actuator bar slide, simulating the plunger in current T34 equipment.  The simulator components at the top simulate the response of the three sensors on the device.  These sensors monitor the insertion of a syringe.  The drop-down selector in front of the plunger sensor simulates the diameter of the syringe inserted into the pump.  The length of the plunger, the diameter of the container tube and the position of the flange can help identify the syringe.  While using the prototype version, the user must interact with the keypad and the sensor checkbox provided in the layout.

Now that the finished prototype is explained correctly and referenced, the in-depth analysis of the code implementation will be more contextual.  The sample code in \ref{code:T34SDA  sample - UI and state } describes the overall instantiation of the State Controller for the System Controller and the User Interface State.  Remember that the System State Controller covers all the system controls for the hardware, and the UI State covers the output interfaces for the user, such as the LCD screen and the state of the lights for alerts.
\begin{figure} 
 \begin{lstlisting}[language= Ada]
pragma SPARK_Mode (On); 
...
-- A program for T34 UI implementation  with Ada-GUI
-- Ada-GUI is based on Gnoga 
-- Copyright (C) by Peterson Jean
--
--  

with Ada_GUI;  
with Control_System;  use Control_System;
with LCD_Display;  
with Simulator;

procedure T34SDA is 

   -- start system variables
   System_State : Control_System.Controller_State;
   Current_UI_State: Control_System.User_Interface_State;
   begin -- T34SDA
   ...
    ScreenRow1 := Ada_GUI.New_Background_Text (Row=> 1 , Column => 2,Text => "<h3>T34 Syringe Driver - Control Unit<h3/>",Break_Before => False);
  ...
  
   LCD_Screen(Graphic , Led ,Current_UI_State, (Is_Display_Off => True,others => <>)); 
  ...
  --  Initialise local data and state,assuming battery level from simulator,
   Control_System.Set_Ctrl_State(( 
                    Previous_Ctrl_State => OFF,
                    Current_Ctrl_State=> OFF,
                    HWSensorData => (Is_Battery_Low => False,   
                                     Battery_level => Simulator.Get_Rand_Percentage   ,
                                     Actuactor_Position => Simulator.Get_Rand_Percentage ,
                                     others => <>), 
                    Pump_Version => "FGDFG858GE",
                    others => <>)
                                );
   Control_System.Set_Supported_Syringe; 
   Control_System.Get_Ctrl_State(System_State);
   PlungerProgress.Set_Value(System_State.HWSensorData.Actuactor_Position);
  
  ...
   Ada_GUI.End_GUI;
end T34SDA;

 \end{lstlisting}
 \caption{T34SDA  sample - UI and state initialisation }
 \label{code:T34SDA  sample - UI and state }
 \end{figure}

 As usual, the pragma aspect for SPARK tells the compiler which switches to use to compile and whether to use the SPARK Ada subset.  In this case, line 1 emphasises that the T34SDA package should be checked against the various SPARK assurance levels.  Lines 9 to 12 import all the necessary packages for the project including the \textit{Ada$\_$GUI }library, the\textit{ Control System} , the \textit{LCD Display }and the \textit{Simulator}.Lines 17 and 18 create the state controller and the user interface state, respectively.  As their instances were created later, the state controller is initialised from lines 27 to 39.  The controller itself will cover the various properties initialised here, but note the call to the simulator to get a random battery percentage.  Why is this important, then?  Because it allows the safety requirements defined earlier in the refined model to be checked randomly, but more importantly, it covers both the internal and external safety risks simultaneously.  The external risk is when the employee mistakenly takes the wrong 9V battery because, visually, there is no way to recognise a low-voltage battery unless it is tested or used.  Internal risk: Considering the previous case, will the system detect the battery level warning and behave correctly?  It proves why the simulator is important in addition to the model checking itself in Ada.  Consequently, it reinforces the argument that formal verification does not replace regular software testing and review but strengthens the process and potentially reduces the required test suite.
 
 Another critical aspect of the implementation is tracking the input key event resulting from the keyboard press.  In Figure \ref{code:T34SDA  sample - Input Event Handling }, the primary event handler runs once the GUI components have been successfully rendered, as shown on lines 2 and 3.  The loop continuously waits for any key input while the device runs and filters the event type for the relevant behavioural result.  On line 6, when the user closes the open window page containing the web server page, it automatically closes the underlying program and all associated running events.  It is worth noting that on line 7, the \textit{Get$\_$Ctrl$\_$State} procedure constantly retrieves the current state of the system controller to ensure that the GUI  synchronises with the hardware at all states.  One of the challenges encountered when handling input key events is how to differentiate between a single key press and a long press, as we are simulating the real T34 in our prototype.  In the current T34 device, a one-touch keypress launches the appropriate key event, while a long press is only valid on the info.  Therefore, the left click is considered a single press action, and the double click is a long press for the prototype implementation.

\begin{figure}
 \begin{lstlisting}[language= Ada]
...
All_Events : loop
Event := Ada_GUI.Next_Event; 

if not Event.Timed_Out then
        exit All_Events when Event.Event.Kind = Ada_GUI.Window_Closed;
        Control_System.Get_Ctrl_State(System_State);  -- always retrieve state upon input events
    if Event.Event.Kind = Ada_GUI.Left_Click then
            Ada_GUI.Log( "Log Event: Left Click");
        if Event.Event.ID = ButtonInfo then -- Info 
                    ... -- other omitted input events  
        elsif Event.Event.ID = ButtonPower then -- power button
            Power_Action : declare 
                Control_System.Get_Ctrl_State(System_State);
            if(System_State.Current_Ctrl_State = Control_System.PRELOADING) then 
                Event_Timer  := Ada_GUI.Next_Event (Timeout => 4.0); -- timeout to wait for user to cancel preloading.

                if Event_Timer.Timed_Out then -- if reached here user has not cancel

                    Update_Control(Control_System.IDLE, Control_System.EMPTY); -- move to actuator  states
                    Control_System.Get_UI_State(Current_UI_State);
                    LCD_Screen(Graphic , Led ,Current_UI_State,(Lign2=> 35, others => <>));
                    Control_System.Get_Ctrl_State(System_State);
                    Ada_GUI.Log(  "state is" & Behaviour_State'Image(System_State.Current_Ctrl_State)); -- Expected state : ACTUATOR_ON
                end if; 
            end if;
                    ...
               end Power_Action;
            else
               null;
            end if;
         end if;
      end if;
      if Event.Event.Kind =  Ada_GUI.Double_Click then
         if Event.Event.ID = ButtonInfo then -- Info
            DC_Info_Action : declare 
            ...
        end if;
    end if;
 end loop All_Events; 
...
 \end{lstlisting}
 \caption{T34SDA  sample - Input Event Handling}
 \label{code:T34SDA  sample - Input Event Handling }
 \end{figure}

Consider the case of the power button where the device has started and correctly initialised until it reaches the "\textit{Preloading}". state.  It is always mandatory to check the most current state of the system controller; otherwise, safety requirements will be violated, both in terms of data integrity and inappropriate states, resulting in some of the hazards defined in the Section  \ref{table:T34hazardsandrisk}. During its  \textit{Preloading} the device performs an internal hardware check which is ignored here for the prototype's scope.  Therefore, an assumption is made to simulate the process by incorporating a four-second timer. In lines 16 and 17, during the 4-second timer, the user can cancel the device \textit{Preloading} by pressing the "NO", otherwise the exit behaviour for the tag transition for that state is executed, resulting in the transition to the next state \textit{Actuator$\_$On}. The current state is evaluated for the correct behaviour by passing the previous and current states and the button type to the state controller manager in our system control unit.  The following section discusses in detail the state controller manager.  As usual, after any change in the system state, the updated state of the user interface is retrieved and then used to update the LCD screen and LED light status (lines 21-22), which facilitates the verification of safety requirement 2.1.1.

While the visual cues are essential for monitoring and controlling the device, the diagnosis of the log is essential for an appropriate response to unexpected incidents or hazards for legal and maintenance purposes.  Therefore, all events are associated with a log entry that tracks the behaviour, status, and exceptions detected when the device is powered.  On line 24, a log entry tracks and reports the new system status of the device. Figure \ref{code:Log  sample - Timestamped diagnostic log } is extracted from the running device prototype log.  Lines 3 to 7, two supported syringe type profiles are loaded, BRAUN Omnifix and Teruno. Both are supported syringe brand models. A few lines down, the code tracks the type of key press and the various states of the system after it has been initialised.
\begin{figure}[h]
 \begin{lstlisting}
...
2022-09-26 03:27:37.59 :  0
2022-09-26 03:27:37.59 : BRAUN Omnifix 
2022-09-26 03:27:37.59 :  0
2022-09-26 03:27:37.59 : BRAUN Omnifix
2022-09-26 03:27:37.59 : Teruno
2022-09-26 03:27:37.59 : 
2022-09-26 03:27:37.59 :  0
2022-09-26 03:27:41.57 : Log Event: Left Click
2022-09-26 03:27:41.57 : PREVIOUS STATE is:OFF
2022-09-26 03:27:43.58 : PREVIOUS STATE is:OFF
2022-09-26 03:27:52.64 : PREVIOUS STATE is:IDLE
2022-09-26 03:27:52.64 : PREVIOUS STATE is:PRELOADING
...
 \end{lstlisting}
 \caption{Log  sample - Timestamped diagnostic log}
 \label{code:Log  sample - Timestamped diagnostic log }
 \end{figure}

The system control package is, in fact, the main central unit that serves as an interface between our model and the other components.  It evaluates, validates and updates the states of the system and the user interface.  It is essential to cover it in depth to understand the verification contracts better.  The specification is essential to define the correct contract or support guidance that the verifier will later use to establish the correctness of the device model. In Figure \ref{code:Control System Code Sample Part 1- Specification  }, several lists are defined to represent the basic types of information to reinforce the code type system and pass the initial phase of the assurance level.  Enumerations are created for all possible states of the T34 prototype and all possible input keys.  On line 6, an "EMPTY" button adds and represents an abstract type for consistency between our refinement model and the code implementation in cases where the label transition does not expect any key input from the user to change its current states.  Like the other list types, the state of the LED can be in one of the enumerated types throughout the execution of the system.  This record type allows SPARK to quickly verify the first level of assurance for security requirement 3.2.1.  The Percentage type protects against overflow errors because we are dealing with integer (number) values.  It helps the prover to check later that the percentage of the battery remains within the allowed limits to avoid side effects. 

\begin{figure}[h] 
 \begin{lstlisting}[language= ada]
package Control_System with SPARK_Mode => On is 
     
   type Behaviour_State is (OFF, IDLE, PRELOADING, ACTUATOR_ON,  SYRINGE_LOADED,SYRINGE_VERIFIED,SYRINGE_CONFIRMED, PUMP_PAUSED, INFUSION_STARTED,INFUSION_STOPPED,INFO);
   -- Behaviour_State defines all different states of the refinement model
   
   type Input_Button_Type is (INFO,UP, DOWN,YES_START, NO_STOP,FF,BACK,ON_OFF,EMPTY);
   -- All available button input from the device interface
   
   type Light_Status is (RED,GREEN,OFF);
   -- All possible color status of LED
   type HW_Sensor is (CLAMP, PLUNGER,FLANGE);
   -- All syringe sensor type
   subtype Percentage is Integer range 0 .. 100;
    
   -- All hardware component status to be passed to main controller state
   type Hardware_State is record 
      Is_Battery_Low : Boolean := False;
      Battery_level: Percentage;
      Is_Barrel_Clamp_Ok: Boolean := False; --sensor data
      Is_Plunger_Ok:Boolean := False; --sensor data
      Is_Barrel_Flange_Ok: Boolean := False; --sensor data
      Actuactor_Position: Natural := 0;
      Occlusion: Natural := 730;
      Max_Rate: Positive :=5;
   end record;
  ...
 \end{lstlisting}
 \caption{Control System Code Sample  Part 1- Specification}
 \label{code:Control System Code Sample Part 1- Specification  }
 \end{figure}
 
  \begin{figure}[h]
 \begin{lstlisting}[language= ada]
   ...
   -- All status of running system as updated by the control system prior to update the UI
   type Controller_State is record
      Previous_Ctrl_State:Behaviour_State;
      Current_Ctrl_State:Behaviour_State;
      HWSensorData: Hardware_State;
      Keypad_Lock: Boolean := False ;
      Pump_Id:String(1 .. 12) := "Syringe Pump";
      Pump_Version: String(1 .. 10):= "----------";
      Pump_Supported_Syringe: Integer;
   end record;
   --All verified status to be passed to UI components of Ada_GUI 
   type User_Interface_State is record
      Primary_Msg:  Ada.Strings.Unbounded.Unbounded_String ;
      Secondary_Msg: Ada.Strings.Unbounded.Unbounded_String ;
      Instructions: Ada.Strings.Unbounded.Unbounded_String;
      Light: Light_Status :=  OFF;  
   end record;
  
   System_State: Controller_State;
   
   function Is_Syringe_Inserted  return Boolean;
   procedure Update_Control (  Previous_State : in Behaviour_State; Button : in Input_Button_Type) 
     with
       Post => System_State.Previous_Ctrl_State = System_State.Current_Ctrl_State'Old;
end Control_System;
 \end{lstlisting}
 \caption{Control System Code Sample Part2 - Specification}
 \label{code:Control System Code Sample Part2- Specification  }
 \end{figure}
 The record type \textit{Hardware$\_$State} defines the hardware entity that represents the current state of the hardware components at a particular behavioural state of the model.  As mentioned earlier, the Controller$\_$State and UI$\_$State of the system are the two record entities used in SPARK Ada to translate the state of our model into two abstract states covering the overall state of the system and the output interfaces at a particular device state.  It is essential to keep track of the controller's previous and current states, as seen in the T34 state machine model  \ref{fig:state diagram}. Some device behaviour states may result in multiple actions on the user interface state depending on the input and its previous behaviour states.  For example, when transitioning to the  "INFUSION$\_$STARTED" state , if the previous state were  $s_6$ would display the new message  "\textit{Pump Delivering}". In contrast, for $s_8 $ the LCD is updated with the same message but generated by a different input trigger. In the requirements on UI and Alerts table section (\ref{table:T34safetyrequirementsUserInterfaces},\ref{table:T34safetyrequirementsAlertAlarm}) ,the requirements 3.2.1 and 2.1.1 can be verified by checking that the abstract User Interface state of the model implementation has a snapshot of the LCD and LED at the particular  Behaviour$\_$State. The \textit{UI$\_$State} can be used to assist the verifier. 
\begin{figure}[!]
\begin{subfigure}{.5\textwidth}
  \centering
  \includegraphics[width=.8\linewidth]{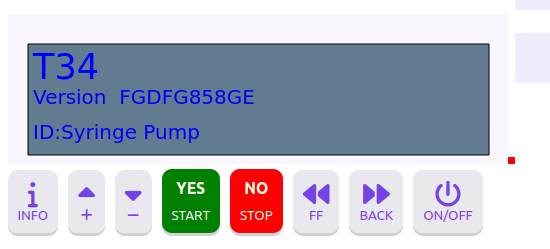}
  \caption{UI$\_$State 1}
  \label{fig:sfig1}
\end{subfigure}%
\begin{subfigure}{.5\textwidth}
  \centering
  \includegraphics[width=.8\linewidth]{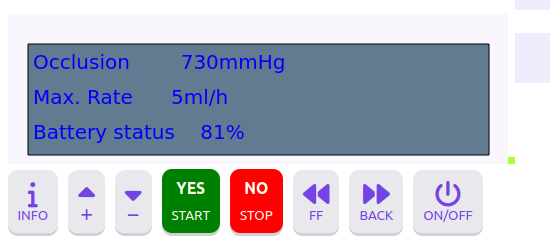}
  \caption{UI$\_$State 2}
  \label{fig:sfig2}
\end{subfigure}
\begin{subfigure}{.5\textwidth}
  \centering
  \includegraphics[width=.8\linewidth]{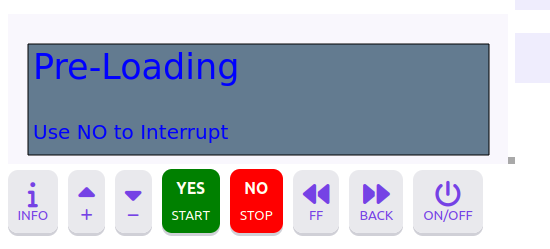}
  \caption{UI$\_$State 3}
  \label{fig:sfig3}
\end{subfigure}%
\begin{subfigure}{.5\textwidth}
  \centering
  \includegraphics[width=.8\linewidth]{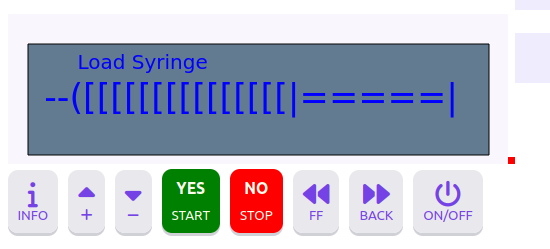}
  \caption{UI$\_$State 4}
  \label{fig:sfig4}
\end{subfigure}
\caption{Prototype: LCD Display Info changes per UI$\_$State}
\label{fig:Changes on UI info}
\end{figure}
 Typically the LCD screen displays its data in 3 main lines. For consistency and to reduce confusing information, the \textit{UI$\_$State} (Lines 15-17) tracks each line as a separate and independent verifiable string. The first two lines are used to issue messages, while the third represents instructions to the user.  The body of the  \textit{Control$\_$System} package  covers most aspects of this rational decision in detail.
 
 The three states of the LED light indicator are visible in the UI prototype  figure \ref{fig:Changes on UI info}. The font size clearly and prominently highlighted the predominant instructions to emphasise these instructions. In \ref{fig:Changes on UI info}.a,\ref{fig:Changes on UI info}.c,\ref{fig:Changes on UI info}.d, the priority is on line 1, line 1 and line 2, respectively. For the UI  state at \ref{fig:Changes on UI info}.c,\ref{fig:Changes on UI info}.d, the unimportant lines 2 and 3 are deleted, respectively, to avoid confusion.
 
 The procedure \textit{Update$\_$Control} in figure \ref{code:Control System Code Sample Part 1- Body  } controls and synchronises the System State and the User Interface State. It accepts two input parameters for the Previous system state and the key input that triggered the State transition.  On line 4, a local time is added to the procedure.  It handles cases where its internal behaviour needs to be delayed or is waiting for some external input from the user via the prototype keyboard.  In line 7, as usual, the procedure registers its state for diagnosis using the internal  Ada$\_$GUI internal logging module. Then the controller evaluates the system's current state and performs the appropriate action for the state transition. 
\begin{figure}[!] 
 \begin{lstlisting}[language = ada]
pragma SPARK_Mode (On);
...
   procedure Update_Control (  Previous_State : in Behaviour_State; Button : in Input_Button_Type) is
      Event_Timer : Next_Result_Info;
      --  Block_Event: Boolean := False;
   begin
      Log ("PREVIOUS STATE is:" &  Behaviour_State'Image(System_State.Previous_Ctrl_State ) );
      case System_State.Current_Ctrl_State is --getting the current state from global controller state
            when PRELOADING =>
         if Previous_State = IDLE and Button = NO_STOP then
            UI_State := (  Ada.Strings.Unbounded.To_Unbounded_String("Max. Rate     " & Positive'Image( System_State.HWSensorData.Max_Rate) & "ml/h"),
                           Ada.Strings.Unbounded.To_Unbounded_String("Occlusion       " & Natural'Image(System_State.HWSensorData.Occlusion) & "mmHg" ) ,
                           Ada.Strings.Unbounded.To_Unbounded_String("Battery status   " & Integer'Image( System_State.HWSensorData.Battery_level )& "%"),
                           others => <>);
            Set_Cur_State(IDLE); 
 ...
 \end{lstlisting}
 \caption{Control System Code Sample  Part 1- Body}
 \label{code:Control System Code Sample Part 1- Body  }
 \end{figure}
 \begin{figure}[!] 
 \begin{lstlisting}[language = ada] 
... 
         elsif Previous_State = IDLE and Button = EMPTY and System_State.HWSensorData.Actuactor_Position = 100 then -- no button pressed and plunger has moved to max positon
            UI_State := (  Ada.Strings.Unbounded.To_Unbounded_String(" --([[[[[[[[[[[[[[[|=====|"),
                           Ada.Strings.Unbounded.To_Unbounded_String( "       Load Syringe   " ) ,
                           Ada.Strings.Unbounded.To_Unbounded_String(""),
                           Light => RED);
            Set_Cur_State(ACTUATOR_ON);
         else  -- plunger in a lower position  use button to move
            Set_Cur_State(ACTUATOR_ON);
         end if;
      end case;
   end Update_Control;
 ...
 \end{lstlisting}
 \caption{Control System Code Sample  Part 2- Body}
 \label{code:Control System Code Sample Part 2- Body  }
 \end{figure}

%----------------------------------------------------------------------------------------
%	SECTION 2
%----------------------------------------------------------------------------------------

\section{Verification  Contracts }

When running the GNATProve flow analysis command, "Examine All" checks and verifies the program with the Bronze Assurance level.  An error is spotted for an exception introduced in the code, which validates the syringe list when initialising the support in the local system memory, in other words, our temporary database.  The argument here is that exceptions are permitted and legit in Ada as with any other programming, whereas SPARK prohibited their use, hence the reported error by the analyser.  If the software is built appropriately to prevent side effects, adding such a mechanism should not be necessary as it invalidates the idea of the safe typeset.  Also, checking in for such errors at runtime is very costly; hence, the rationale is always to establish ways to prove them never to happen instead of raising them.  To solve this, the author replaces the exception within the branching condition for the index of the list of syringes (Appendix \ref{AppendixB}, Figure \ref{code:Log  sample - Adding Syringe procedure }). In the full code implementation in Appendix \ref{AppendixB}, Figure \ref{code:Code  - Syringe Package post-verification  }, the log code is moved to the if statement checking the number of indexes found by the Search procedure parsing the syringes profiles list. When the index exceeds 0, a duplicate was found, which does not require us to save the new syringe profile.  Running GNATprove's verifier on the Syringe package verified that no SPARK violation was detected for the exception.  

\begin{figure}[h]
 \begin{lstlisting}
syringe.adb:47:07: error: handler is not allowed in SPARK
syringe.adb:47:07: error:violation of pragma SPARK_Mode at line1
 \end{lstlisting}
 \caption{GNATProve Log - Syringe package dataflow}
 \label{code:Log  sample - GNATprove syringe package  log }
 \end{figure}

 Now GNATProve successfully runs its prover on the Syringe package but reported multiple warnings and information regarding the passed checks.In the GNATProve log (figure \ref{code:Log GNATprove syringe package missing contract } ) one  high warning for the \textit{Last$\_$Index} in the "\textit{Add}" procedure is reported . It should be addressed to pass the data dependencies checks.  Therefore we got some suggestions from the prover on ways to possible address the missing dependency in the specification contract for the Add procedure. The "\textit{Depends}" clause in the contract of figure \ref{code:Log  Add procedure's verified   specifications }indicates that \textit{Last$\_$Index }of the found syringe profile in the syringe list is global variable access inside the Add procedure block.  Therefore SPARK requires that it is explicitly defined as data dependencies for the \textit{Add} procedure. Such an addition helped the prover verify the missing checks and ensure the safety requirements. 
\begin{figure}[!]
 \begin{lstlisting}
syringe.adb:41:10: high: "Last_Index" must be listed in the Depends aspect of "Add" (SPARK RM 6.1.4(14))[#1]
syringe.ads:28:06: info: flow dependencies proved[#0]
syringe.ads:36:08: info: flow dependencies proved[#2]
syringe.ads:37:15: medium: postcondition might fail[#11]
 \end{lstlisting}
 \caption{GNATProve Log - Syringe package missing contracts}
 \label{code:Log GNATprove syringe package missing contract }
 \end{figure}

\begin{figure}
 \begin{lstlisting}[language= ada]
procedure Add (List: in out Syringe_List; Syringe: Syringe_Type )
     with
       Pre => Last_Index < 15,
       Depends => (List => (List, Syringe,Last_Index) , Last_Index => (Last_Index,List, Syringe)),
       Post => (if (Last_Index in List'Range )then  List(Last_Index) = Syringe); 
 \end{lstlisting}
 \caption{Code Sample - Add procedure's verified   specifications  }
 \label{code:Log  Add procedure's verified   specifications }
 \end{figure}

Although the prototype runs and compiles correctly, the last few reports have demonstrated the need for a prover.  The rate function used to calculate the pump's flow rate across the 24-hour administration can be seen in figure \ref{code:Basic Flow Rate specification and Proofs }. Running GNATprove on the rate function outputs that the float overflow check is essential to show that the return type is respected between volume divided by the hour. This output is possible as the hour has to be set in its float type, 24.0 instead of 24.  It is not yet sufficient for SPARK to conclude that all proofs are checked and proven for the rate.  Thus the divide by zero warning on line 12, along with the hint on how to help the prover.

  \begin{figure}[h]
 \begin{lstlisting}[language=ada]
 function Rate(This: Syringe_Type) return Real ;

function Rate(This:  Syringe_Type) return Real is
   begin
      return This.Volumes(ml)/24.0;
   end Rate;

----------------------------------------------------------------
Phase 1 of 2: generation of Global contracts ...
Phase 2 of 2: analysis of data and information flow ...
syringe.adb:8:30: info: float overflow check proved[#3]
syringe.adb:8:30: medium: divide by zero might fail [possible fix: subprogram at syringe.ads:27 should mention This in a precondition][#4]
 \end{lstlisting}
 \caption{Basic Flow Rate specification and Proofs}
 \label{code:Basic Flow Rate specification and Proofs }
 \end{figure}

   \begin{figure}[h]
 \begin{lstlisting}[language=ada]
 function Rate(This: Syringe_Type) return Real with
     Depends => (  Rate'Result => This),
     Pre=> This.Volumes(ml) > 0.0,
     Post => Rate'Result > 0.0 and  Rate'Result <= 5.0 ;
 
----------------------------------------------------------------
Phase 1 of 2: generation of Global contracts ...
Phase 2 of 2: flow analysis and proof ...
syringe.adb:8:30: info: float overflow check proved[#4]
syringe.adb:8:30: info: division check proved[#5]
 \end{lstlisting}
 \caption{Flow Rate verified with Proofs}
 \label{code:Flow Rate verified with Proofs}
 \end{figure}
 
% \begin{figure}[h]
%  \begin{lstlisting}
% phase 1 of 2: generation of Global contracts ...
% Phase 2 of 2: analysis of data and information flow ...
% lcd_display.adb:32:07: error: "Set_New_Widget" is not allowed in SPARK (due to plain precondition on dispatching subprogram)
% lcd_display.adb:32:07: error: violation of pragma SPARK_Mode at line 1
% lcd_display.adb:33:07: error: "Set_Background_Color" is not allowed in SPARK (due to plain precondition on dispatching subprogram) 
% ...
% lcd_display.adb:61:07: error: violation of pragma SPARK_Mode at line 1
% control_system.ads:64:29: info: initialization of "State" proved[#0]
% control_system.ads:72:27: info: initialization of "State" proved[#1]
% gnatprove: error during analysis of data and information flow
 
%  \end{lstlisting}
%  \caption{Log  sample - Timestamped diagnostic log}
%  \label{code:Log  sample - Timestamped diagnostic log }
%  \end{figure}

% \begin{figure}
%  \begin{lstlisting}
% function Rate(This:  Syringe_Type) return Real is
%   begin
%       return This.Volumes(ml)/24.0;
%   end Rate;
% -----------------------------------
% syringe.adb:8:30: info: float overflow check proved[#3]
% syringe.adb:8:30: medium: divide by zero might fail [possible fix: subprogram at syringe.ads:27 should mention This in a precondition][#4]
%  \end{lstlisting}
%  \caption{GNATProve Log - Syringe package dataflow}
%  \label{code:Log  sample - GNATprove syringe package  log }
%  \end{figure}

% Chapter Template

\chapter{Results and Discussion} % Main chapter title

\label{Chapter4} % Change X to a consecutive number; for referencing this chapter elsewhere, use \ref{ChapterX}

%----------------------------------------------------------------------------------------
%	SECTION 1
%----------------------------------------------------------------------------------------
\section{Overview}
Throughout this research, the focus has been on safety, from assessing the existing body of research on verifying medical devices to their interfaces.  Such evaluation has been explored in two ways and progressively integrated into this research design.  It began with the creation of an appropriate model from the GIP that abstracts and encompasses the safety of reverse engineering for the T34 syringe pump.  This model is reviewed during the Design Verification, where some design errors, such as in the user interfaces, are manually corrected.  They are fixed according to the safety requirements defined for the T34.  This step mainly covers the qualitative aspect of verification.  The model is implemented correctly in Ada 2012, which incorporates great language features to avoid some side effects.   The second part of the work focused on ensuring that the level of assurance is increased as much as possible to the highest level, which is platinum, through verification conditions, as seen in the previous section.  It covered one major package, "\textit{syringe}", dealing with the profile of the syringe recorded during the device's initialisation.  For example, this package is essential for the User Interface as its data is used to update the user interface when calculating the flow rate or other critical information such as the brand name of the syringe.

More importantly, the code implemented for this project has great potential for educational purposes.  Overall it is an opportunity to test the non-binding Graphical UI library in Ada, which the choice of Ada$\_$GUI has justified.  It extends the library's potential from the traditional approach of demonstrating algorithms by developing a more complex and realistic example that developers may be better able to develop and adapt. 
 
\section{Effort consideration}
Following the work to build the adaptive refinement model of the T34 case study, most of the time was spent on setting up the GUI library as isolated from the control system as possible to reflect the exact abstraction deduced in the model.  In the model, the design choice was to isolate all user interface code from the system code, which was done by segmentation into packages.  Extending the Ada$\_$GUI to work correctly with this difference created packages such as LCD$\_$Display and the control$\_$system  representing the model's user interface and system controls.  A new procedure had to be added to the core  Ada$\_$GUI package to allow  the LCD$\_$Display to replace and recreate the canvas representing the screen in the prototype.  This replacement was necessary because the user interface stacked the generated canvas on top of each other.  The correction was, therefore, effective in simulating behaviour similar to that of an actual device screen. 

Verification is the other major block in this project.  It is not tedious because adapting the verification to achieve many GNATprove proof checks sometimes requires changing or refactoring some aspects or logic of the code.  Handling loops and calculations require more checks for proof because sometimes, the deeper or more nested the branching, the more difficult it is for GNATprove to find the correct checks.  Therefore, contracts must be modified gradually in such cases, as GNATprove proposes new checks to pass.  Sometimes GNATprove also can mislead in case of some bad defined contracts.  Hence the need to spend more time and effort on defining the correct formal specification for each subprogram and updating the annotations to help the prover automatically find the proofs. 

\section{Proofs Analysis}
A qualitative analysis of this research highlights the success of the proposed design model when refined.  It allows for the detection of some generic safety-based design errors, which supports the general idea motivated in the background search to have a more standardised and manufacturer-independent resource framework.  The role of such a framework is to force the manufacturer to anticipate other hazards that may not yet be relevant to their devices but may be relevant to similar devices.  A successful example of the potential of the refinement model is the error found in the existing T34 user interface when displaying critical information and instructions to the user, following safety requirement 2.1.1 on user interfaces.  The discovery was that there was confusion in the information displayed, which was corrected using the refined security requirements of the GIP project.   The hazard, in this case, was how the confusion between an instruction and a user message could lead to potential harm to a patient by delaying the nurse's decision while validating a syringe already recognised by the device.  The message "\textit{Check syringe}" contradicts the instruction "\textit{Loaded correctly}".  As seen in the figure in the Device User Interface Design section, the nurse can potentially interpret the "Check Syringe" message as an instruction to reinsert the syringe when the message is passive information from the previous step of the device.  The message "\textit{Check Syringe}" was proposed as it is more explicit than "\textit{Check Syringe}"

The implementation improved the refined T34 model by using the well-defined abstraction subroutine.  Such a structure made it easy to see which unit layer of the abstraction model is signalled by GNATprove.  It thus makes it easier to deal with missing SPARK annotations by component or sub-program and provides better readability and traceability.  GNATprove provides quantitative data on the number of proof checks and tests as a function of the level of proof. In the table \ref{tab:Assurance Level of proof reached} the assurance level is presented for each of the base packages representing the refined T34 model. Through the analysis of GNATprove, the author concluded that each package had reached the Silver level. In Section \ref{Adacore's spark framework},  the level of proof \textit{Silver} means that the prototype subprogram or package has successfully proven the absence of runtime errors.  In addition, it implicitly implies validation of the flow analysis, proving that data and flow dependencies have also been proven for much of the code in compliance with the SPARK Ada subset.

Ada$\_$GUI is ignored because it is added as a library; therefore, errors resulting from GNATprove regarding SPARK violation were ignored because it is implemented in Ada 2012.  Although it uses contracts such as Pre and Post conditions, it does not enforce any formal specification of the SPARK subset pragma.  Therefore, the error for violating the SPARK pragma in the \textit{LCD$\_$Display} is ignored.  For the package to achieve a higher level of assurance as "\textit{Gold}", the use of Bounded String must be enforced according to the UI security requirements on the information.  Strings will have to be limited and checked to a maximum length of 15 characters so that they still fit on a small screen and do not wrap around on a new line or get cut off.  For the control system, the contracts must be updated to ensure that the states at the start of the system calls are always checked as well as when the system control unit exists.  The syringe, as most details in the previous section, shows the rationality of the verification requirements and the process needed from one level to another, e.g. verification of the add procedure for the syringe size list, proving that the security requirements on data integrity and availability are met.  As a critical component, verifying syringes was essential to ensure that the user could select the correct pre-loaded syringe and that the automatic loading of syringes was done correctly.  Overall, the level achieved is satisfactory for the prototype, and quantitative analysis could be carried out when higher proof levels are achieved, as this level is still insufficient to claim full proof.

\begin{table}[]
    \centering
    \begin{tabular}{||l l l ||}  
 \hline
 \textbf{Component/Package} & \textbf{Level of Assurance }& \textbf{Steps to Higher Proof}\\ [0.5ex] 
 \hline\hline
Syringe & Silver & Loop$\_$Invariant  improvement \\
LCD$\_$Display & Silver & Bounded String \\ 
  Control System & Silver &  Pre and Post Conditions\\[1ex] 
 \hline  
\end{tabular} 
 \caption{Results -  Assurance Level of proof reached}
 \label{tab:Assurance Level of proof reached}
\end{table} 
  
% Effort needed for development
% reflection on the level of assurance reached 
% INput key challenge . Long Press from Key stuck. therefore can't verify for the safety requirements for key stuck due to limitation of the event action

% limitations verifying each line of lcd display easier but more difficult to implement non ASCII value or value that can't be represented as string. as we verified the string size and length. Limitations on bounded string.
% design mistake: error message to user "check syringe" able to find based on the generic safety requirements.
%-----------------------------------
%	SUBSECTION 1
%-----------------------------------
\section{Summary}
The results obtained with the prototype are satisfactory and demonstrate the model's effectiveness in verifying the T34 user interface.  However, other SPARK annotations could be revised to verify further and achieve higher assurance. 
% Chapter Template

\chapter{Conclusion} % Main chapter title

\label{Chapter5} % Change X to a consecutive number; for referencing this chapter elsewhere, use \ref{ChapterX}

%----------------------------------------------------------------------------------------
%	SECTION 1
%----------------------------------------------------------------------------------------
  
\section{General Overview}
 
Safety-critical systems are today ubiquitous in people's lives.  Their critical role has many benefits and risks.  Many examples exist, but the most contextual is the application of medical devices to patients.  As benefits, one can count the lives saved in the hospital.  At the same time, this benefit is associated with the high risk of these types of devices.  The more these devices are adopted for many patients, the more defects are detected.  A typical device in this category is an infusion pump.  Many reported incidents are related to user interface design errors that have gone unnoticed.  In other cases, they result from confusion about the user interface display.  Therefore, there is an urgent need to constantly update existing research as the number of new devices increases each year dramatically.

%-----------------------------------
%	SUBSECTION 1
%-----------------------------------
 
 \section{Research Significance}
 In this research, the author has explored the use of formal methods to verify systems.  More importantly, it covers the potential role of formal methods in typical and traditional software development concerning the industrial style of device development processes.  This research also explores how formal verification can be used to reduce the overheads in testing and review that are costly in resources and expertise.  With the development of the automatic prover, formal verification is becoming increasingly available to achieve this goal.  Reviewing existing research on the use of formal methods led to the adoption of generic infusion pump resources such as risk, state controller and design faults in developing the proposed T34 syringe driver model.  The T34 syringe pump, due to its portability and wide use in ambulatory care, is a perfect case study to evaluate the possibility of such methodologies.  SPARK Ada, a strong type subset of the Ada language, provided a verification tool like GNATprove to assist the developer in the tedious task of proving the safety of the T34 code and model.  Such a task demonstrated the use of Spark Ada in user interface verification for safety-critical medical devices.  The contribution of this work is the portability of safety requirements for generic infusion pumps using Spark Ada with refinement methods.  At the end of this project, a reliable prototype based on a realistic behavioural model of the T34 infusion pump has been developed.  The initial objective is to establish a practical case study for industrial simulation or even the implementation of real devices.  More importantly, it contributes to the existing body of knowledge on implementing GIP.
\section{Limitations}
The developed prototype offers much potential for further exploration and promises insight into a 100\% complete Ada source code without binding code. One of the challenges in achieving this goal is that the SPARK subset of Ada adds and removes certain features of the Ada language.  Implementing a full SPARK is currently limited and may sometimes require tricking the static verifier by passing True as an assertion that tends to prove a particular requirement by forcing its warning to be ignored.  Another challenge is the limitation of the GUI library, which is not yet fully ported to SPARK as some code relies on unsupported SPARK Ada code.  The GUI libraries are either in the development phase or adopted by the community as binding libraries.  For example, the simulation of specific events, such as a long press versus a stuck key in a GUI, was not evident in \textit{Ada$\_$GUI}.  For this project, the author assumed that no keys were jammed to eliminate some complexity and avoid introducing a bug into the GUI library.  
\section{Overall Conclusions}
Finally, future work could be directed towards supporting and developing a GUI library such as Ada$\_$GUI, which could extend the portability of SPARK to full stack development.  This work contribution would implicitly affect the portability of the proposed prototype, as it could increase the assurance of a more secure and portable tool that can be used for training and real-world simulation.  It will also be interesting to make a comparative study for the GIP model by comparing the number of faults or design errors detected by two formal verification tools like PVS and SPARK on the same device model.  The comparison will further argue the success of the Generic Infusion Pump and contribute to a formalisation of its use by industries.  This research, therefore, concludes with the importance of user interface verification in improving safety assurance and its potential for actual industrial development.  
% \include{Chapters/Chapter6} 

%----------------------------------------------------------------------------------------
%	THESIS CONTENT - APPENDICES
%----------------------------------------------------------------------------------------

\appendix % Cue to tell LaTeX that the following "chapters" are Appendices

% Include the appendices of the thesis as separate files from the Appendices folder
% Uncomment the lines as you write the Appendices

% Appendix A

\chapter{Additional Documents} % Main appendix title

\label{AppendixA} % For referencing this appendix elsewhere, use \ref{AppendixA}

\section{Other Resources}

\begin{itemize}
\item Latex template by:
Vel (vel@latextemplates.com)\cite{noauthor_latex_nodate}
\item Initial report from  author's previous works via Swansea University canvas.
\item Project Specification from  author's previous works via Swansea University canvas.
\end{itemize}

\begin{itemize}
\item 
\end{itemize}
 
\section{Hazards and Risk Analysis - GIP Model  \cite{zhang_hazard_2010}}
 
\begin{figure}[h]
\centering
\includegraphics[scale=0.4]{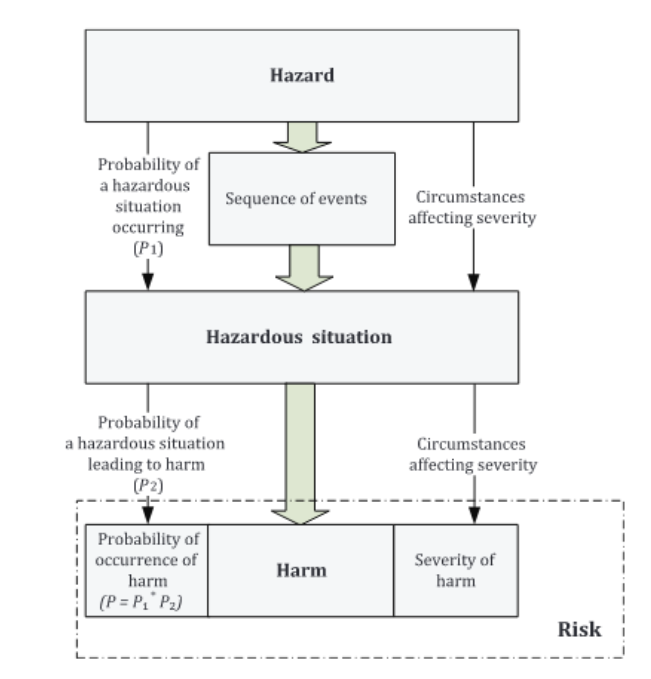}
\decoRule
\caption{Flow - risk analysis from hazards   }
\label{fig:hazardoussituations}
\end{figure}

\subsection{Hazardous Situations \cite{zhang_hazard_2010}}
\begin{figure}[h]
\centering
\includegraphics[scale=0.4]{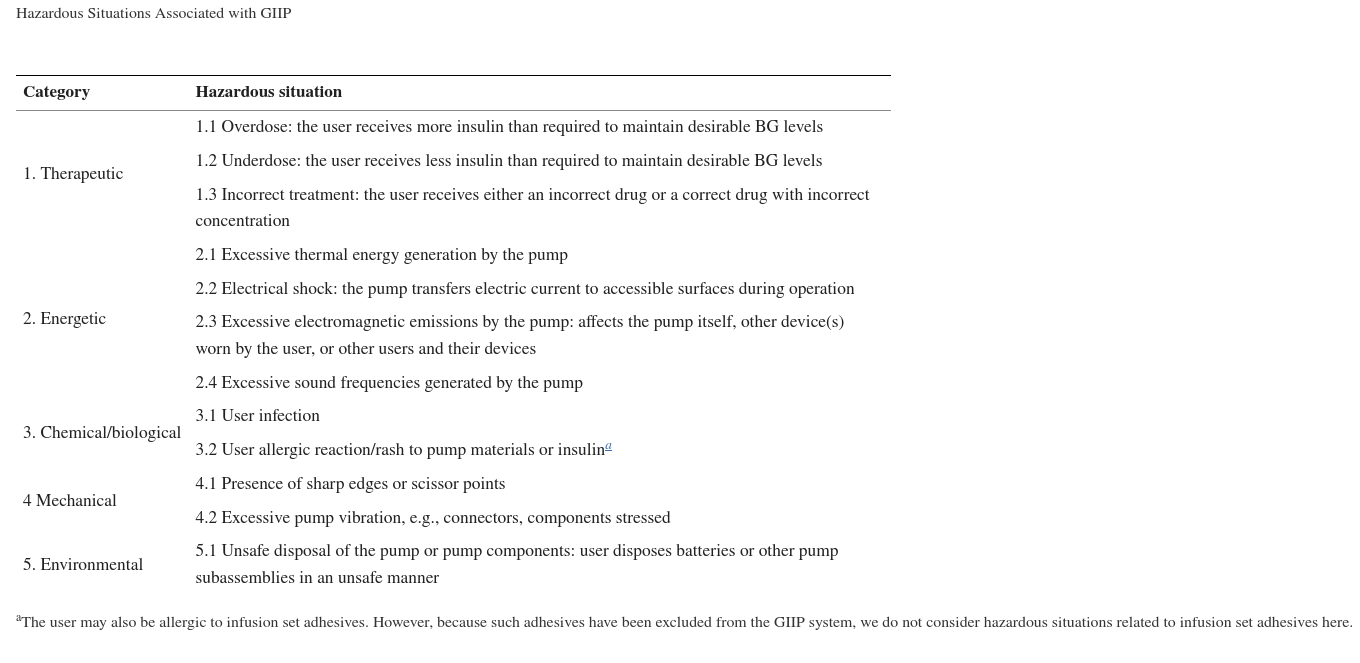}
\decoRule
% \caption{  }
\label{Appendix:hazardoussituations}
\end{figure}

\subsubsection{Operational Sources of Hazardous Situations}
\begin{figure}
\centering
\includegraphics[scale=0.4]{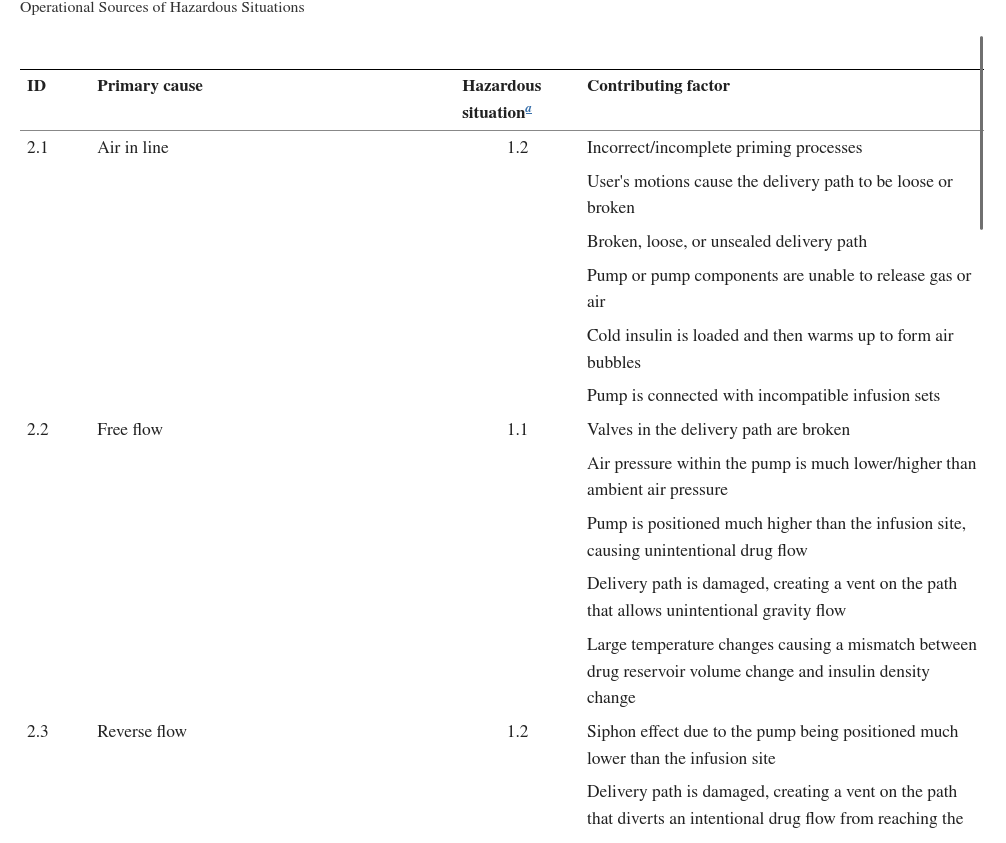}
\includegraphics[scale=0.4]{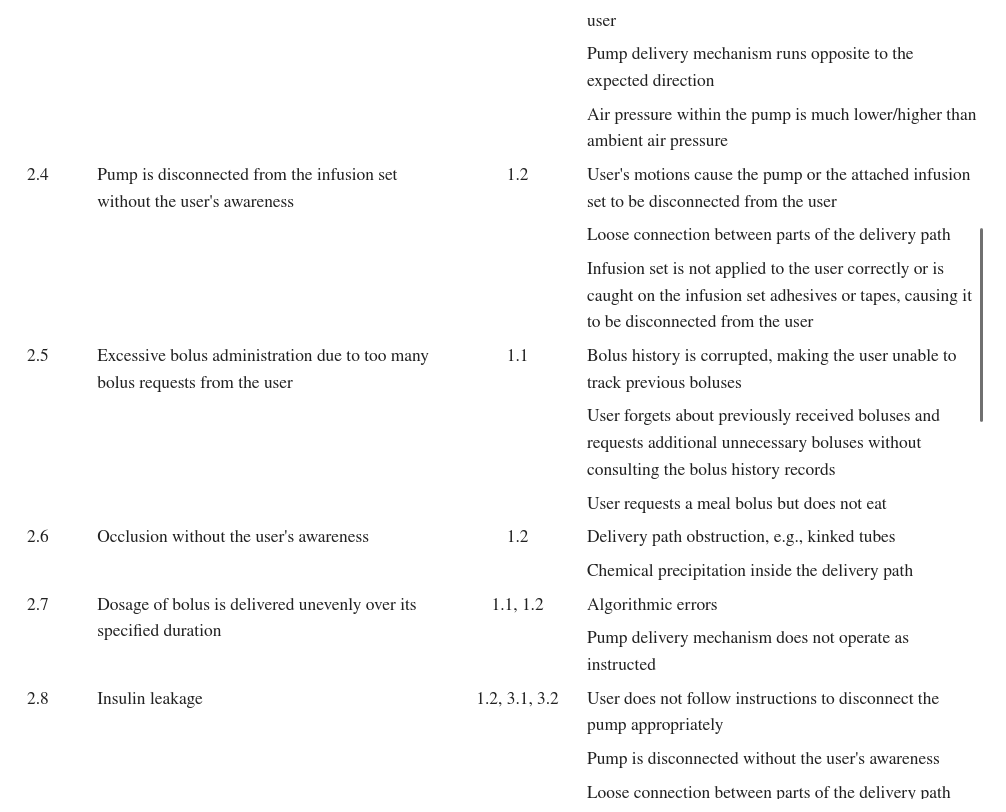}
\includegraphics[scale=0.4]{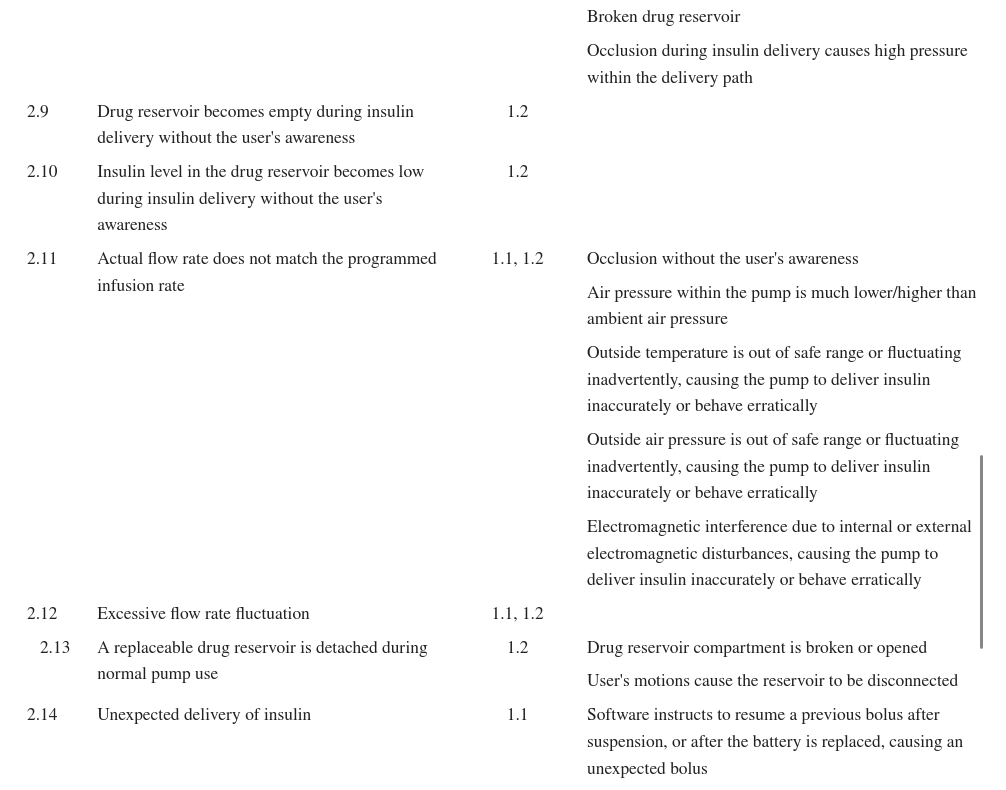}
\includegraphics[scale=0.4]{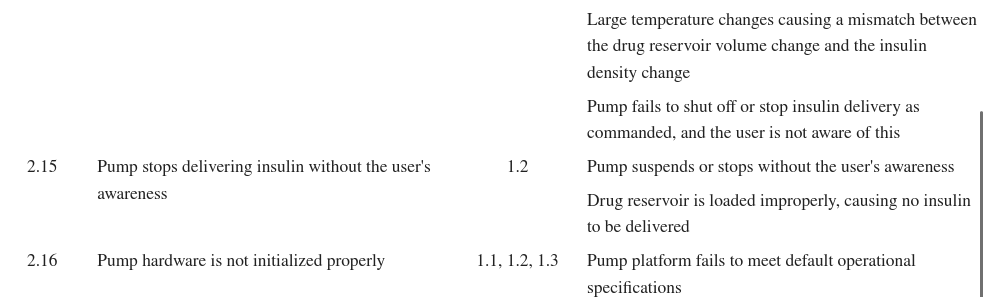}
\decoRule
% \caption{  }
\label{Appendix:sourceshazardoussituations}
\end{figure}

\subsection{Full T34 Refinement model}
\begin{figure}[h]
\centering
\includegraphics[scale=0.3]{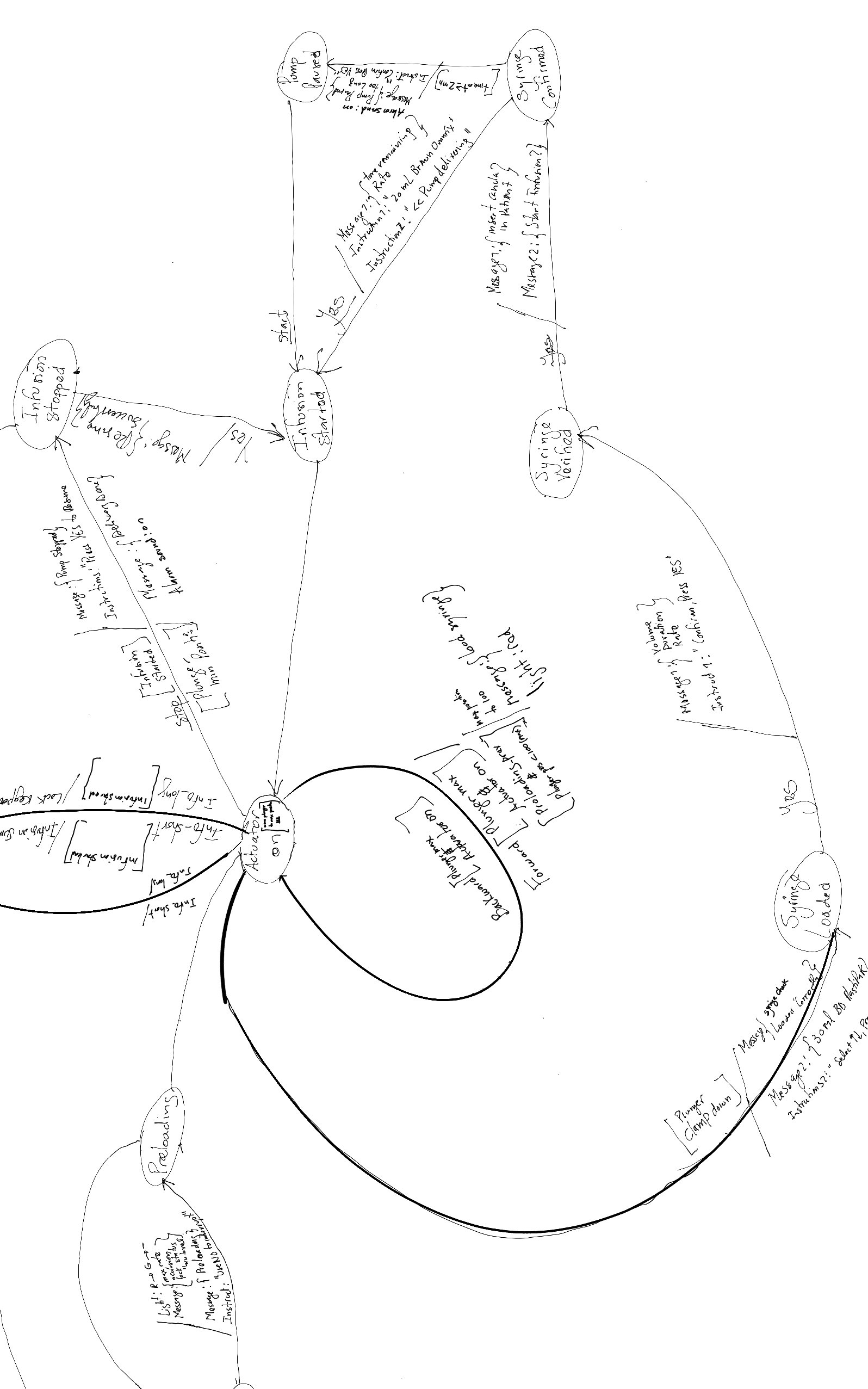}
\decoRule
\caption{  }
\label{Appendix:hazardoussituations}
\end{figure}

% The color of links can be changed to your liking using:

% {\small\verb!\hypersetup{urlcolor=red}!}, or

% {\small\verb!\hypersetup{citecolor=green}!}, or

% {\small\verb!\hypersetup{allcolor=blue}!}.

% \noindent If you want to completely hide the links, you can use:

% {\small\verb!\hypersetup{allcolors=.}!}, or even better: 

% {\small\verb!\hypersetup{hidelinks}!}.

% \noindent If you want to have obvious links in the PDF but not the printed text, use:

% {\small\verb!\hypersetup{colorlinks=false}!}.

\chapter{Additional Source Code } % Main appendix title

\label{AppendixB} % For referencing this appendix elsewhere, use \ref{AppendixA}
 
\begin{figure}[h]
 \begin{lstlisting}[language=ada]
 procedure Add (List: in out Syringe_List; Syringe: Syringe_Type ) is
      Index: constant Integer := Search(List, Syringe);
     
   begin
      if Index >0 then 
         raise EXISTING_DUPLICATED_ERROR;
            
      else
         Insert(List,Syringe); 
      end if;
      
   exception
      when EXISTING_DUPLICATED_ERROR =>
         Ada_GUI.Log("Error: Duplicate");
   end Add;
 \end{lstlisting}
 \caption{Code Sample Log - Adding Syringe procedure with Exception }
 \label{code:Log  sample - Adding Syringe procedure  }
 \end{figure}

\begin{figure}[h]
 \begin{lstlisting}[language=ada]
  
pragma SPARK_Mode (On); 
with Ada_GUI; 
with Ada.Strings.Unbounded; use Ada.Strings.Unbounded;
package body Syringe is

   function Rate(This:  Syringe_Type) return Real is
   begin
      return This.Volumes(ml)/24.0;
   end Rate;
   
   function Search (List:  Syringe_List; Syringe: Syringe_Type ) return Integer is 
      count: Natural :=0;
   begin
      
      for item1 of List loop
         Ada_GUI.Log(To_String(item1.Brand));
         count := (if To_String(item1.Brand) = To_String(Syringe.Brand) and Syringe.Volumes(ml) = item1.Volumes(ml) then count +1 else count);
          
    
      end loop;
      Ada_GUI.Log("" & Integer'Image(count));
      return count;
   end Search;
   
   procedure Insert (List: in out Syringe_List; Syringe: Syringe_Type ) is
   begin
      Last_Index := Last_Index + 1;
      List(Last_Index) := Syringe;
      
   end Insert;
   
   procedure Add (List: in out Syringe_List; Syringe: Syringe_Type ) is
      Index: constant Integer := Search(List, Syringe);
     
   begin
      if Index >0 then 
         Ada_GUI.Log("Error: Duplicate Syringe Preset");
         --therefore no need for the raised exception , the duplicate syringe type is ignored and not saved
         -- and appropriate log event saved. 
      else
         Insert(List,Syringe); 
      end if;
       
   end Add;
end Syringe;


 \end{lstlisting}
 \caption{Code  - Syringe Package post-verification  }
 \label{code:Code  - Syringe Package post-verification  }
 \end{figure}

\FloatBarrier

\section{Installation Manual}
\begin{enumerate}
\item GNAT is the compiler use for development and running the prototype. Go to GNAT's website  \href{ https://www.gnu.org/software/gnat/}{here} to download it or alternatively skip to  \textit{  step 2}. 
\item Adacore's website \href{https://www.adacore.com/download}{here} to download thee GNAT Studio Communitiy Edition  which has a default compiler and all the necesary tools.
\item Open your terminal and navigate to the codesource repository  \begin{lstlisting}
cd ~/pathtorepository/T34DriverPrototypeSource 
 \end{lstlisting}
\item Compile the \textit{Ada$\_$GUi} dependency
 \begin{lstlisting}
gnatmake ada_gui.adb
 \end{lstlisting} 
 After successfull completion you can open the project now in GNATStudio
\item  From the terminal you go to the project T34$\_$Syringe$\_$Driver$\_$Ada folder then open \textit{gnatstudio } to launch it or use your favourite launcher
\begin{lstlisting}
cd ../T34_Syringe_Driver_Ada 
 \end{lstlisting}
\item Now in GNATStudio to build and run the prototype  , Go to \textit{Build > Run  t34sda }. It should launch the prototype with the following screen
\item  For full use details , please see the NHS guide here \cite{nhs_education_for_scotland_guidelines_nodate}
\begin{figure}[h]
\centering
\includegraphics[scale=0.4]{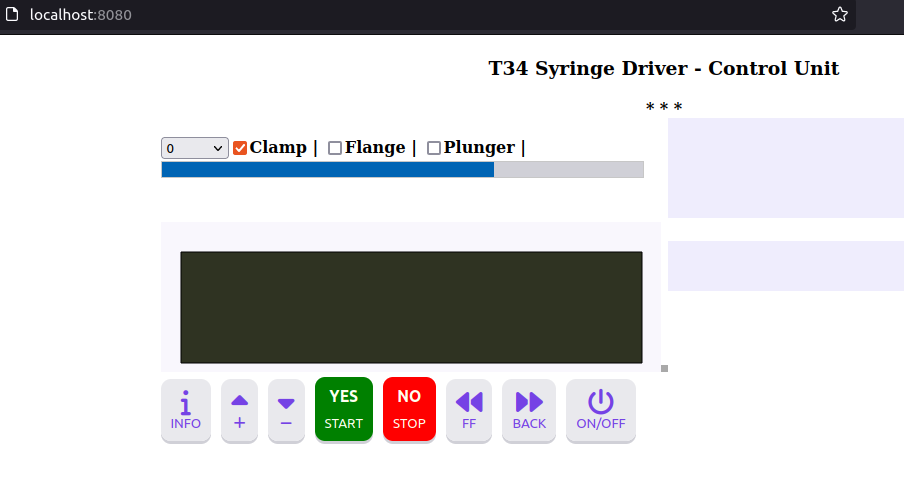}
\decoRule
\caption{Welcome page prototype  }
\label{fig:T34}
\end{figure}

\end{enumerate}

%----------------------------------------------------------------------------------------
%	BIBLIOGRAPHY
%----------------------------------------------------------------------------------------

% \printbibliography[heading=bibintoc]

\bibliography{main}

\begin{thebibliography}{10}
\bibitem{nhs_education_for_scotland_guidelines_nodate}
{\sc for Scotland, N.~E.}
\newblock Guidelines for the use of the {T34TM} {Ambulatory} {Syringe} {Pump}
  by {CME} {Medical} for adults in palliative care.
\end{thebibliography}
%----------------------------------------------------------------------------------------
% \input{main.bbl}

\end{document}